\documentclass[graybox, secnum]{svmult}
\usepackage{mathptmx}       
\usepackage{helvet}         
\usepackage{makeidx}         
\usepackage{graphicx}        
\usepackage{multicol}        
\usepackage[bottom]{footmisc}
\usepackage{hyperref}        
\usepackage{soul}            
\hypersetup{colorlinks=true,urlcolor=blue,citecolor=cyan}
\usepackage[force]{feynmp-auto}
\usepackage{amsmath,amssymb}

\DeclareMathOperator\arccosh{arccosh}

\def\an[#1,#2]{\left\langle#1\,#2\right\rangle}
\def\aq[#1,#2,#3]{\left\langle#1|#2|#3\right]}
\def\qa[#1,#2,#3]{\left[#1|#2|#3\right\rangle}
\def\sq[#1,#2]{\left[#1\,#2\right]}
\def\spa#1.#2{\left\langle#1\,#2\right\rangle}
\def\spab[#1,#2,#3]{\left\langle#1|#2|#3\right]}
\def\spba[#1,#2,#3]{\left[#1|#2|#3\right\rangle}
\def\spb#1.#2{\left[#1\,#2\right]}
\def\lor#1.#2{\left(#1\,#2\right)}

\usepackage[square,numbers,sort&compress]{natbib}
\def\triangleboxleft{\scalebox{.9}{$\triangleleft$}\kern-.1em\Box}
\def\triangleboxright{\Box\kern-.1em\scalebox{.9}{$\triangleright$}}
\def\dBox{\Box\kern-.1em\Box}

\makeindex             

\begin{document}
\title*{Effective Field Theory and Applications: Weak Field Observables from Scattering Amplitudes in Quantum Field Theory}
\titlerunning{Post-Minkowskian Observables from Quantum Field Theory}
\author{N. Emil J. Bjerrum-Bohr \thanks{corresponding author}, Ludovic Plant\'e and Pierre Vanhove}
\institute{N. Emil J. Bjerrum-Bohr \at Niels Bohr Institute, Blegdamsvej 17, DK-2100 Copenhagen, Denmark, 
\and Ludovic Plant\'e \at Niels Bohr Institute, Blegdamsvej 17, DK-2100 Copenhagen, Denmark 
\and Pierre Vanhove \at Institut de Physique Th\'eorique
Orme des merisiers, CEA/Saclay
F-91191 Gif-sur-Yvette Cedex, France 
}
%
%
\maketitle
%
\abstract{In this chapter, we will review the field-theoretic treatment of General Relativity based on an effective field theory extension of the Einstein-Hilbert action. This pragmatic route to low-energy quantum effects in gravity critically underpins miscellaneous investigations of phenomenological and quantum extensions of General Relativity. We discuss how it allows quantum field theory to be a theoretical laboratory for testing Einstein's theory of gravity and demonstrate the current state of the art of an efficient and practical scheme for evaluating the classical components of perturbative weak-field scattering amplitudes until the fourth post-Minkowskian order. Such results complement numerical predictions in Einstein's theory of gravity.}

\section*{Keywords} 
Quantum field theories of gravity, Effective field theory, Classical Einstein gravity, gravity phenomenology, Observables in general relativity

\section{Introduction}
The present cosmological paradigm hinges on General Relativity -- Einstein's theory of gravity -- crafted as a relativistic theory of curved space-time. Gravity is inferred universal -- with dynamics that couple equally to all varieties of matter. This framework successfully conditions our image of the shape and dynamics of the Universe. We introduce dark energy and matter because General Relativity fails to explain the observed dynamics and stability of the galaxies or the accelerated expansion of our Universe, and the tension in the measured value for the acceleration rate of our Universe~\cite{Freedman:2021ahq} \footnote{\footnotesize{{\it E.g.} from local astrophysical data and the cosmic microwave background and anomalies in the current cosmological model~\cite{Peebles:2022akh}}}, bids us assess transformations of the law of gravity over large astrophysical scales. \\[5pt]
Recent cosmological surveys focus on understanding such queries, despite the weakness of the gravitational interactions that make direct detection of departures from Einstein's gravity very challenging.
A new exciting development is observations of gravitational waves from Earth-based interferometers \cite{LIGOScientific:2016aoc} that have lessened the observational gaps in the scales to which we gauge gravity with accuracy. 
This new pursuit aims to shed light on the exact gravitational attraction of massive astrophysical objects such as black holes and neutron stars. Planned space-based gravitational wave interferometers will even bring this a step further and have the capacity to address questions associated with gravitational attraction and the expansion of the Universe. \\[5pt]
We are thus constantly learning more about classical gravitational attraction, but quantum gravity is a critical puzzle in theoretical physics. Everywhere in Nature, we encounter mechanics dynamics, yet a century after the creation of General Relativity by Albert Einstein, a fundamental theory of quantum gravity still lacks an adequate resolution. In place of such a description of gravity -- effective field theory offers a possible attractive avenue for exploring low-energy quantum gravity consequences. \\[5pt]
The overarching focus of this review is to demonstrate how low-energy field theory methods provide a flexible and efficient starting point for quantum field theory applications in gravity. Held up against the geometrical framework of General Relativity -- this is a different way to think about gravity -- a quantum field theory mediated by quanta -- spin-2 gravitons -- that propagate the field and generate metric and curvature in a perturbative expansion. As a field particle, the graviton is massless because, like photons, we give it an infinite range. \\[5pt]
The advantage of this adaption of gravity is that it allows a direct unification with other fundamental forces at low energies in the context of the standard model. For instance, we verify the classical equivalence principle at the microscopic level by considering the scattering of different types of matter in the context of the bending of light around a huge massive star and demonstrate that the classical scattering angle is universal as expected. \\[5pt]
Another point is improving the analysis of the measured gravitational wave signals in observations of binary in-spirals that will be capable of testing Einstein's classical theory \cite{Einstein:1916vd}. Traditional perturbative (off-shell) quantum field theory calculations in gravity are far from optimal computation-wise. Directly computation from Feynman diagrams is notoriously complex and tiresome; far more problematic than for gluon scattering amplitudes in gauge theories as they rely on an infinite number of vertices and endless index contractions. This dire need for progress evokes the search for inspiration in (on-shell) computational methods for particle physics amplitudes, where, prompted by the requirements of the Large-Hadron-Collider advancement is recent. Stimulated by early examples of gravity scattering amplitude computations \cite{Iwasaki:1971vb,Neill:2013wsa,Bjerrum-Bohr:2013bxa} and \cite{Damour:2016gwp,Damour:2017zjx}, a dedicated programs were outlined in \cite{Bjerrum-Bohr:2018xdl,Cheung:2018wkq} and this has been a catalyst for new technology. It exploits that classical physics materializes when quantum numbers are enormous. An excellent application is quantum scattering amplitudes with superheavy black holes as point particles. For black holes with no spin see for instance~\cite{Cristofoli:2019neg,Bern:2019nnu,Antonelli:2019ytb,Bern:2019crd,Parra-Martinez:2020dzs,DiVecchia:2020ymx,Damour:2020tta,DiVecchia:2021ndb,DiVecchia:2021bdo,Herrmann:2021tct,Bjerrum-Bohr:2021vuf,Bjerrum-Bohr:2021din,Damgaard:2021ipf,Brandhuber:2021eyq}. Current state of the art for such computations is~\cite{Bern:2021dqo,Bern:2021yeh,Bern:2022jvn} reaching fourth post-Minkowskian order while the fifth post-Minkowskian order approached in the probe limit~\cite{Bjerrum-Bohr:2021wwt}. While our principal emphasis here will be the non-spinning amplitude-based computations there are also significant improvements for spinning black holes, see e.g.,~\cite{Guevara:2017csg,Vines:2017hyw,Arkani-Hamed:2017jhn,Guevara:2018wpp,Vines:2018gqi,Chung:2018kqs,Guevara:2019fsj,Maybee:2019jus,Arkani-Hamed:2019ymq,Damgaard:2019lfh,Aoude:2020onz,Chung:2020rrz,Bern:2020buy,Haddad:2020tvs,Guevara:2020xjx,Kosmopoulos:2021zoq,Bautista:2021wfy,Haddad:2021znf,Jakobsen:2022fcj} and some based on the world-line
approaches, e.g.,~\cite{Kalin:2020fhe,Mogull:2020sak,Jakobsen:2021smu,Jakobsen:2021zvh,Dlapa:2021npj,Jakobsen:2022fcj}.\\[-15pt]
\begin{figure}[h]
\centering{\includegraphics[width=7cm]{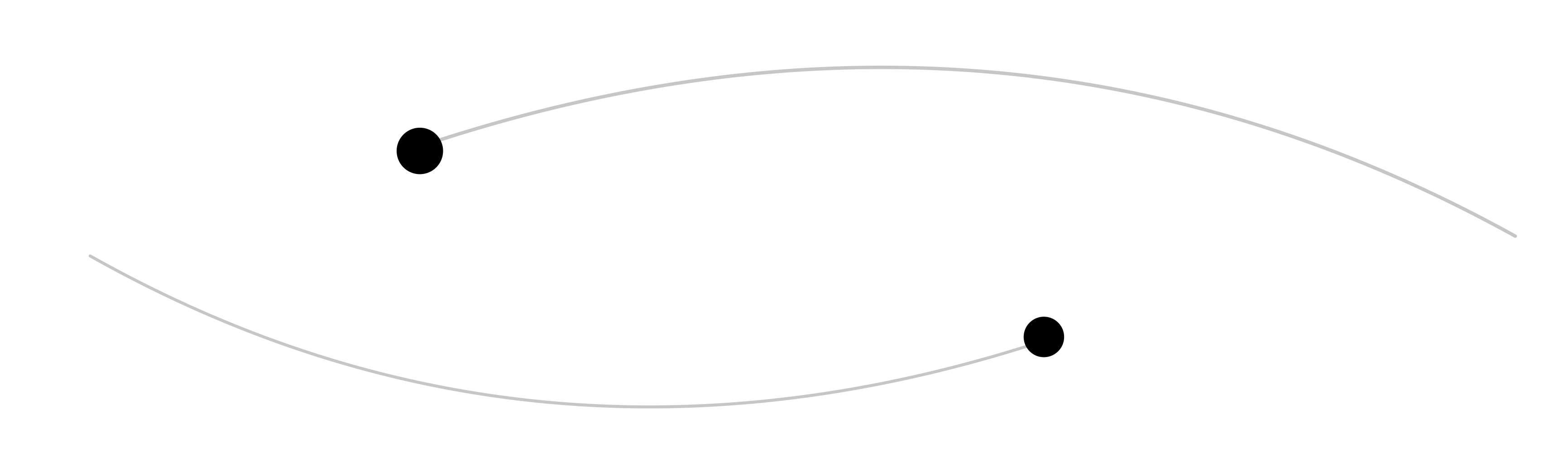}}
\caption{\small Schematic inspiral of two black holes.}
\end{figure}
 \\[5pt]
A key point in such applications is a relationship that connects tree amplitudes for open and closed strings found by Kawai-Lewellen-Tye, -- a relationship dubbed the 'double-copy' \cite{Bern:2008qj} or 'KLT' \cite{Kawai:1985xq}.
\begin{figure}[h]
\centering{\includegraphics[width=5cm]{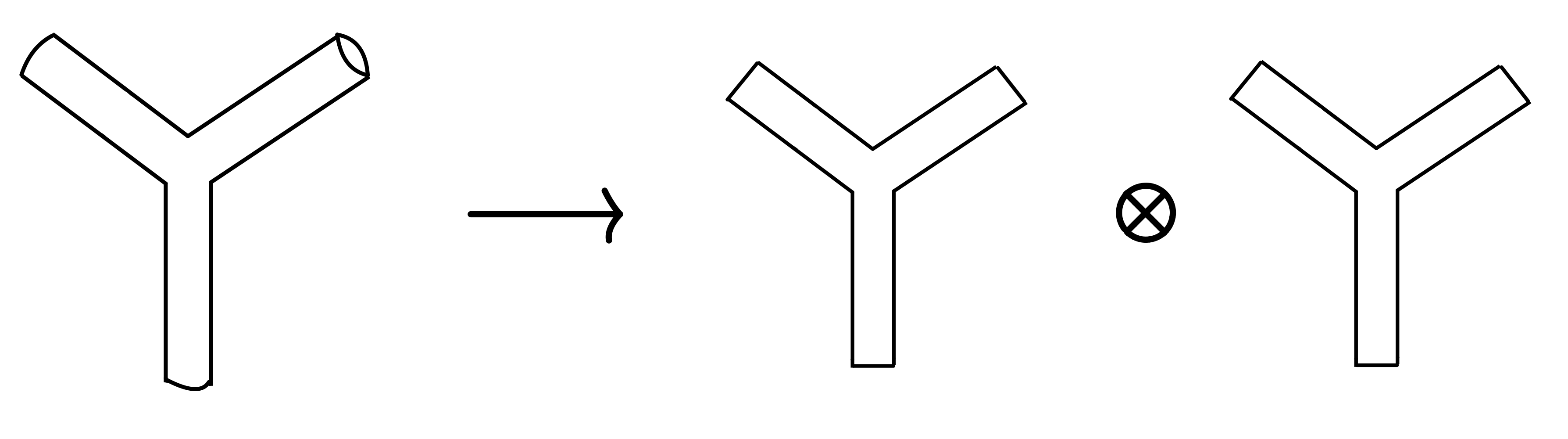}}

{$\downarrow$ (field theory limit) $\downarrow$}

\centering{\includegraphics[width=5cm]{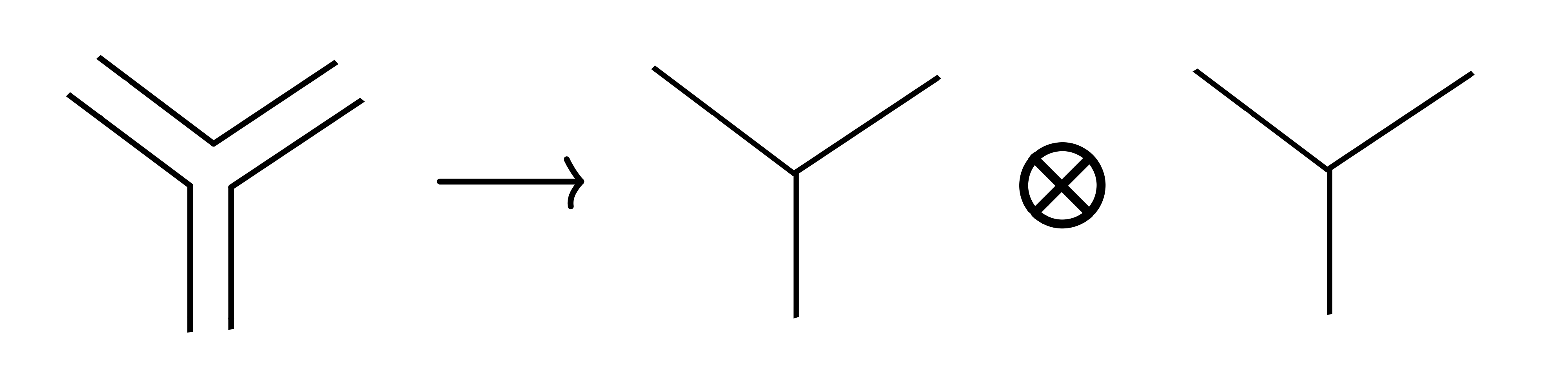}}
\caption{\small Top: schematic relation between a three-point correlation function for the closed string (left) and two open string correlation functions (right). Bottom: schematic relation between a gravity three-point vertex and two three-point gluon vertices induced by string theory.}
\end{figure}
Yang-Mills theory and gravity are very different theories with rather different Lagrangian dynamics -- but double-copy relations provide clear evidence for a new fundamental connection between these forces of Nature. It implies that one can recycle amplitude techniques and provide a possible way to compute amplitude gravity directly from the Yang-Mills theory. It is highly non-trivial and with many intriguing implications that we are still grasping.  \\[5pt]
We will undoubtedly continue to have endless questions regarding gravity and the evolution of the Universe -- we hope that the framework of the effective field theory extension of the Einstein-Hilbert action, combined with new efficient computational techniques, will provide inspiration for the formulation of a fundamental theory of gravity guided by observations. \\[5pt]
We organize the review as follows. First, we will make a lightning review of the effective field theory treatment of General Relativity. We will consider how treating General Relativity as an effective field theory is a natural framework for investigations of gravitational attraction and how the effective
field theory treatment splits the long-range infra-red and
short-distance ultra-violet physics contributions in terms of a
well-defined energy scale expansion. In computations, this is helpful, as we can neglect short-distance physics compared to long-range physics at the length scales probed by observations. Next, we will introduce a framework for computations and consider how the reinstatement $\hslash$ in momenta to helpful for separating classical and quantum effects. We will compute gravitational scattering amplitudes using generalized unitarity and consider how to infer observables and metrics in General Relativity from amplitudes. Finally, we will conclude and look ahead.

The following chapters, to appear in this volume, are related to the
present discussion, by presenting an effective field field theory
point of view from  the  compact binary system
dynamics~\cite{Goldberger:2022rqf} and general discussion of quantum
effects in general relativity from an effective field theory point of view~\cite{Donoghue:2022eay}.

\section{Lightning introduction of the effective field theory treatment of General Relativity}
Treating General Relativity as an effective field theory is a natural framework for explorations of gravitational attraction in the context of quantum field theory. Historically it has been understood since the 1960ties that Einstein's general theory based on the Einstein-Hilbert action relativity has a perturbation expansion in terms of gravitational scattering amplitudes~\cite{Feynman:1963ax,DeWitt:1967yk,DeWitt:1967ub,DeWitt:1967uc}. Starting from a path integral and with minimal assumptions, Feynman derived the Einstein-Hilbert action regarding gravity as a particle theory.
\begin{equation}\label{e:HE}
{\cal L}=\sqrt{-g}\Bigg[\frac{R}{16 \pi G_N}\Bigg].
\end{equation} 
It is customary to define a weak field expansion around the flat Minkowski space-time metric $\eta_{\mu\nu}$ by  
$g_{\mu\nu}(x)\equiv \eta_{\mu\nu} + \sqrt{32 \pi G_N}h_{\mu\nu}(x)\,,$
and utilize this and the definition of the Riemann curvature tensor 
$R^\mu_{\
\nu\alpha\beta} \equiv
\partial_\alpha \Gamma_{\nu\beta}^\mu
-\partial_\beta \Gamma_{\nu\alpha}^\mu
+\Gamma^\mu_{\sigma\alpha}\Gamma_{\nu\beta}^\sigma
-\Gamma_{\sigma\beta}^\mu\Gamma_{\nu\alpha}^\sigma,$ 
as well as the determinant of the metric field $\det(g_{\mu\nu})$ to 
expand the theory in powers of $G_N$~\cite{Veltman:1975vx}. Following a gauge-fixing procedure, one derives the following propagator in harmonic gauge as
\begin{equation}
\displaystyle \frac {i{\cal
P}^{\alpha\beta,\gamma\delta}}{q^2+i\varepsilon}\,,
\quad
{\cal P}_{\alpha\beta;\gamma\delta} =
\frac12\left[\eta_{\alpha\gamma}\eta_{\beta\delta} +
\eta_{\beta\gamma}\eta_{\alpha\delta}
-\eta_{\alpha\beta}\eta_{\gamma\delta}\right]\,.
\end{equation}
Expressions for three-point-graviton vertices are derived in~\cite{DeWitt:1967yk,DeWitt:1967ub,DeWitt:1967uc,Sannan:1986tz} and with background field theory vertices in~\cite{Bjerrum-Bohr:2002ks,Bjerrum-Bohr:2002gqz,Bjerrum-Bohr:2014lea}. Ghost can be accounted for by the Fadeev-Popov approach.\\[5pt]
In this manner, one can regard perturbative computations in gravity as in traditional particle physics. Nevertheless, such a quantization steers into ultra-violet renormalization issues due to the dimensionful gravitational coupling $G_N$. This was demonstrated explicitly in 1974 by 't Hooft and Veltman~\cite{tHooft:1974toh} who directly evaluated the one-loop ultra-violet divergence and uncovered the following result
\begin{equation}
{\cal L}_{\rm UV\ divergence} = \frac{1}{8 \pi^2}\frac{1}{D-4}\Bigg[\frac{ 1}{120} R^2+ \frac{7}{120} R^{\mu\nu}R_{\mu\nu} \Bigg]\,.
\end{equation}
Here $D$ represents the space-time dimension. At each order in perturbation theory, a new divergence emerges causing the theory to be non-renormalizable, and the resolution in the context of effective field theory is to assume that the Hilbert-Einstein term in~\eqref{e:HE} is the leading term in a Lagrangian expansion that includes every conceivable generally
covariant function of the metric and its derivatives. We develop the effective
Lagrangian in the invariants ordered in the magnitude of their derivative contributions. This can be formally understood~\cite{Weinberg} as the expansion obtained from integrating out heavy mass degrees of freedom in favor of light in a fundamental theory below a specific scale. We thus introduce new higher-derivative couplings in terms of an infinite series of effective operators and this innately leads~\cite{Donoghue:1993eb} to the treatment of General Relativity as an effective field theory.
\begin{equation}\label{e:Leff}
{\cal L}_{\rm eff}=\sqrt{-g}\Bigg[\frac{R}{16 \pi G_N}+ C_1 R^2 + C_2 R^{\mu\nu}R_{\mu\nu}+\ldots\Bigg]\,.
\end{equation}
From a low-energy effective viewpoint, all the uncertainties about
the ultra-violet completion of the theory are retained in
the coefficients $C_i$, which have to be specified by experiments
and observations. As a classical theory, it is essential to recognize
that the above Lagrangian describes a theory of General Relativity that
is augmented by new operators even as a classical theory. When
considering effective actions there is the question of the magnitudes
of the coefficients in the derivative expansion
in~\eqref{e:Leff}. Weinberg explained that any theory of massless
spin-2 reduced to Einstein's gravity at large distances~\cite{Weinberg:1965rz}, and it has been asserted in~\cite{Camanho:2014apa,Caron-Huot:2022jli} that the coefficients are parametrically
suppressed by some higher-spin scalar $M$ to be specified. \\[5pt]
The coefficient $C_1$ of the Ricci scalar squared term
in~\eqref{e:Leff} is dimensionless. From measurements of the absence of
derivation from the $1/r$ law, the E\"ot-Wash collaboration~\cite{Adelberger:2003zx}
inferred that $C_1$ could take very large value $0\leq C_1< 10^{61}$. The
conceivable presence of such a large coefficient in the effective
action could be alarming, but this is not the case if it is regarded in the context of an
$f(R)$ theory of gravity, where the $C_1$ decides that the mass of the
scalaron $m^2=8\pi G_N/C_1$~\cite{Stelle:1977ry}. For a value $C_1\sim 10^{61}$ the
scalaron has a mass of the order $10^{-3}eV/c^2$ like the neutrinos~\cite{Brax:2019iut}.\\[5pt]
It should be noted that gravitational wave observations open an
intriguing quest to bound such coefficients even further by comparing
observation to theory in particular in the context of multi-messenger events~\cite{TheLIGOScientific:2016src,Barausse:2020rsu}. Here the effective field theory treatment must separate the long-range infra-red and
short-distance ultra-violet physics contributions permitting the extraction of
long-range, low-energy results in the theory that is independent of
the high-energy behavior of the theory. \\[5pt]
\subsection{Matter couplings}
Including matter in the effective treatment of General Relativity can be accomplished considering the Lagrangian
\begin{equation}
{\cal L}={\cal L}_{\rm eff}+ {\cal L}_{\rm matter}\,,
\end{equation}
where the effective gravitational Lagrangian~\eqref{e:Leff} is minimally coupled to the matter field $\sqrt{-g}{\cal L}_{\rm matter}=g^{\mu\nu}T_{\mu\nu}$. This includes in a covariant way matter couplings in the theory.
The rationale for following this route is that the gravitational interaction between two Schwarzschild black holes are
obtained from gravitational scattering amplitudes between scalar
fields minimally coupled to gravity by their stress-energy tensor 
\begin{equation}
T_{\mu\nu} \equiv \partial_\mu \phi\, \partial_\nu \phi 
- \frac{\eta_{\mu\nu}}{2}\left(\partial^\rho\phi\partial_\rho\phi-m^2 \phi^2\right)\,,\quad {\rm with}\quad
\sqrt{-g}{\cal L}_{\rm matter}\equiv g^{\mu\nu} T_{\mu\nu}\,.
\end{equation}
The multipole expansion has been provided at the first post-Minkowskian order
in~\cite{Vines:2017hyw} and the second post-Minkowskian order for aligned spins with the orbital angular momentum~\cite{Vines:2018gqi}. The multipole expansion from the spin of the Kerr black hole demands
external states with spin. In the case of spin
$S=0,1/2,1$~\cite{Bjerrum-Bohr:2002ks,Holstein:2008sw,Holstein:2008sx}, spin
$S=2$~\cite{Vaidya:2014kza} have been computed at the one-loop level. Since external states of spin, $s$ lead to
a multipole expansion to the order $2s$ one must consider as well
massive external states of higher spin~\cite{Guevara:2017csg,
 Vines:2017hyw, Arkani-Hamed:2017jhn, Guevara:2018wpp,
 Guevara:2019fsj}.
While for photons and massless fermions we derive analogous stress-energy tensors, see for instance \cite{Bjerrum-Bohr:2014zsa} for further details. Following the analysis accomplished using the techniques utilized in
\cite{Donoghue:1993eb,Bjerrum-Bohr:2002aqa,Bjerrum-Bohr:2002ks,Bjerrum-Bohr:2002gqz}, where the following scalar vertices are uncovered in $p_1$ in and $p_1'$ out-going convention.
The two-scalar-one-graviton vertex $ \tau_1^{\mu
    \nu}(p_1,p_1')$ is
\begin{equation}\label{e:tau1}
\tau_1^{\mu\nu}(p_1,p_1') = \frac{i\kappa_{(4)}}2\left[p_1^\mu p_1'^{\nu}
+p_1^\nu p_1'^{\mu} -\frac12 \eta^{\mu\nu}\,(p_1+p_1')^2\right]\,.
\end{equation}
The two-scalar-two-graviton vertex $\displaystyle
  \tau_2^{\eta\lambda\rho\sigma}(p_1,p_1')$ is
\begin{eqnarray}\label{e:tau2}
\tau_2^{\eta \lambda \rho \sigma}(p_1,p_1')&=&- {i\kappa_{(4)}^2} \bigg [ \left
\{{\cal P}^{\eta
\lambda, \alpha \delta} {\cal P}^{\rho \sigma,\beta}_{\ \ \ \ \delta} +
\frac14\left\{\eta^{\eta \lambda} {\cal P}^{\rho \sigma,\alpha \beta} +
 \eta^{\rho \sigma} {\cal P}^{\eta \lambda,\alpha \beta} \right \}
\right \} \cr && \left (p_{1\,\alpha} p_{1\,\beta}' + p_{1\,\alpha}' p_{1\,\beta}
\right ) 
+\frac14 {\cal P}^{\eta \lambda,\rho \sigma} \,( p_1+ p_1')^2\bigg]\,.
\end{eqnarray}
We evaluate all contributing diagrams at four points and compute the tree level two-scalar-two-graviton amplitude as follows.\\[-25pt]
\begin{figure}[hhht]
  \centering
  \includegraphics[width=2.8cm]{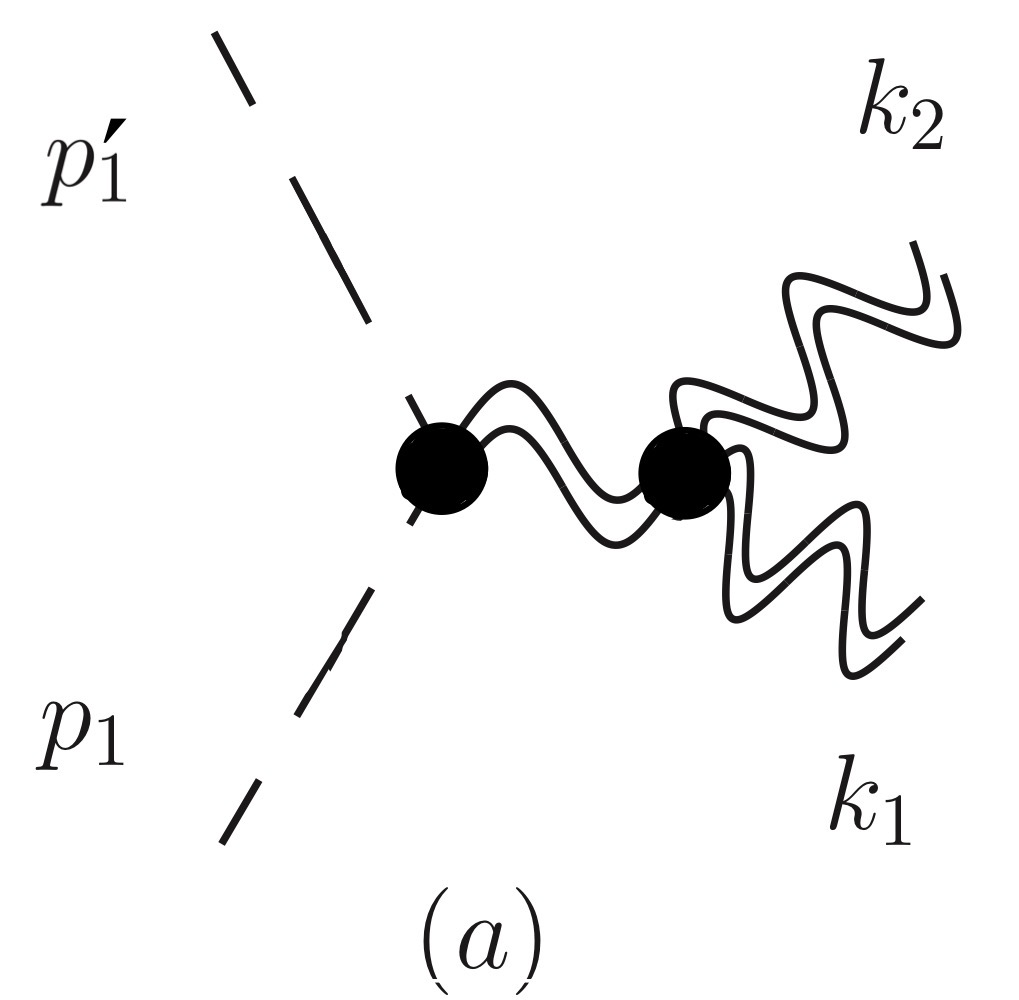}
  \includegraphics[width=2.5cm]{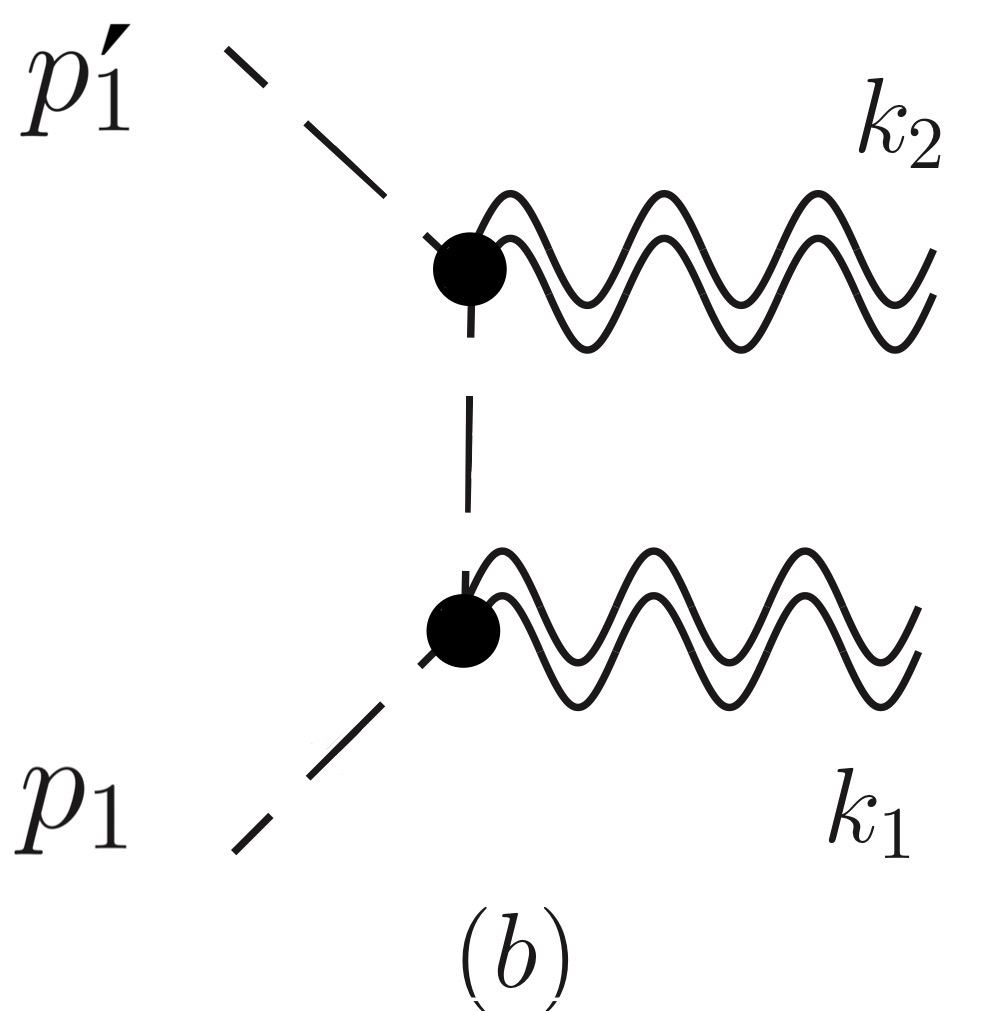}
  \includegraphics[width=2.7cm]{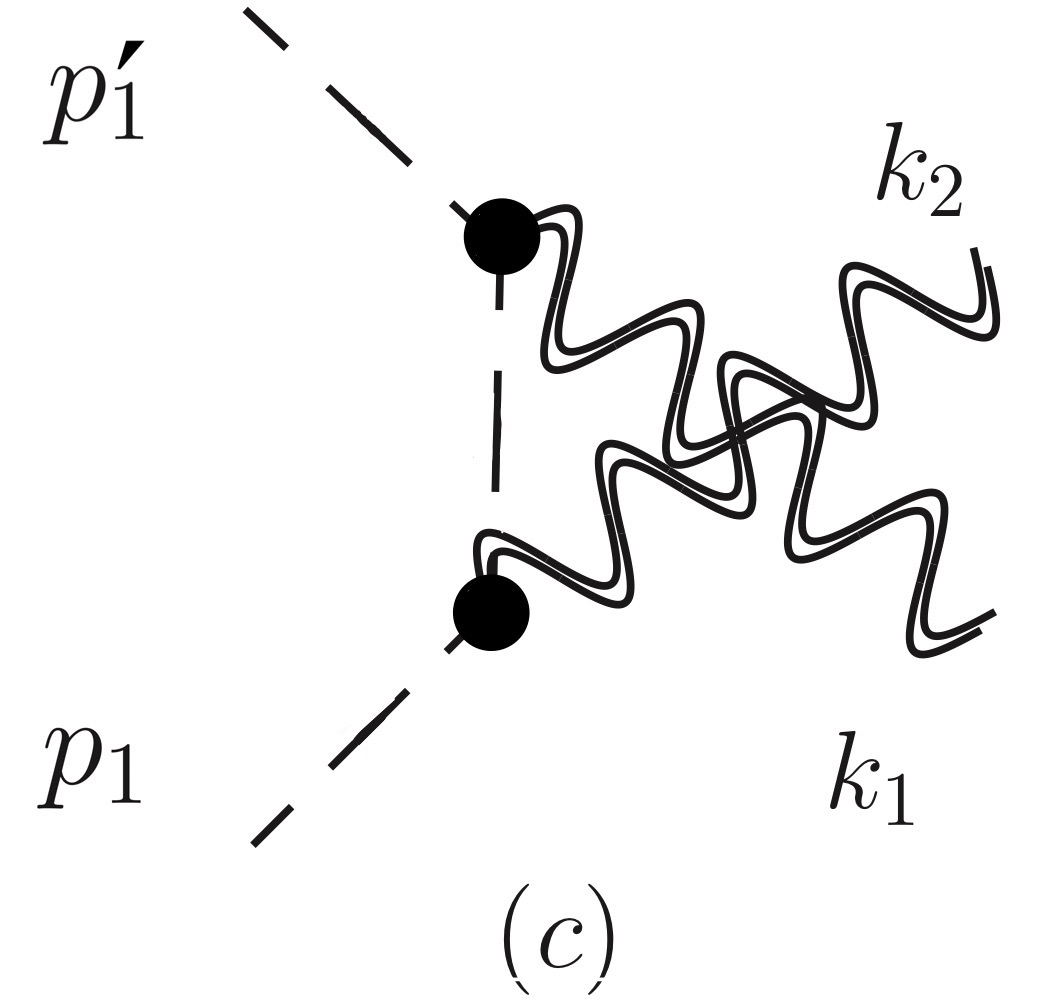}
  \includegraphics[width=2.5cm]{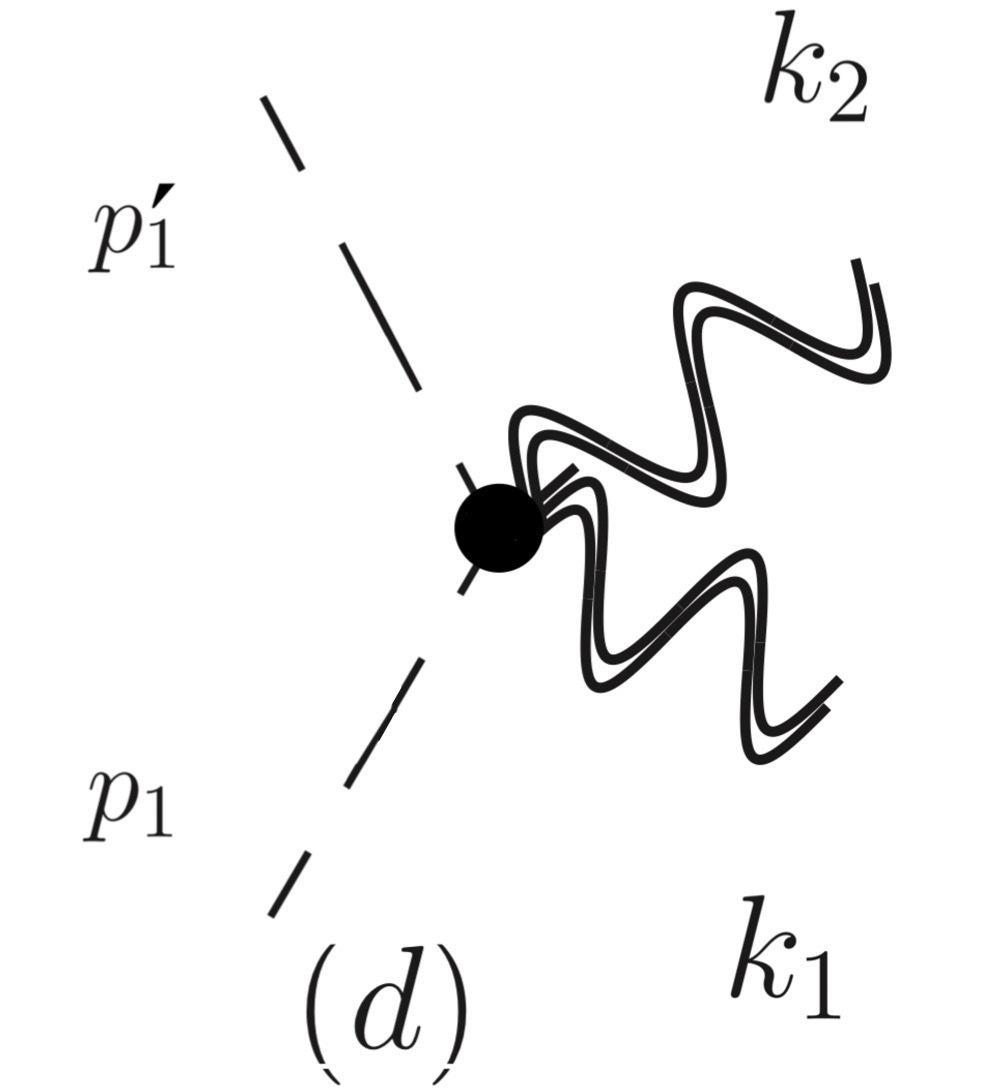}
  \caption{\small The various contributions, (a) $s$-channel, (b) $t$-channel, (c)
    $u$-channel, (d) contact term.}
  \label{fig:treeemission}
\end{figure}
\section{Gravity amplitudes from quantum field theory}
As we have witnessed in the last section, we can regard perturbative computations of gravitational scattering amplitudes in the context of effective field theory. Early treatments of gravity relied on unwieldy off-shell vertex rules and lead to somewhat unmanageable expressions due to both the immense number of permutations and the considerable index contractions. This immense growth in complexity renders amplitude computations in pure gravity theory to be exceptionally problematic. Thus it is advantageous that we have developed strategies in recent years where unitarity integrated with more painless on-shell remedies for tree amplitudes in gravity supplies pathways to streamline computations significantly.

The benefit of using the spinor-helicity variable for external legs is more compact results for amplitudes. We will in this section outline such computations and the associated formalism. Generating amplitudes from Feynman graphs one arrives at expressions with terms consisting of monomials of 
\begin{equation}
 (k_i \cdot k_j)\,,\quad (\zeta_i \cdot \zeta_j)\,,\quad  (\zeta_i \cdot k_j)\,,
\end{equation}
where $k_i^\mu$ is the momentum of the $i$th incoming leg and $\zeta_i^\mu$ the associated polarisation vector. In principle, the polarisation vectors $\zeta_i^\mu$ can be arbitrary, but gauge theories we have learned that it leads to incredibly compact expressions to utilize spinor-helicity variables. In this formalism, we compute all amplitude configurations with polarisation vectors for positive and negative helicity gluons defined by
\begin{equation}\begin{split}
(\zeta_i^+)_\mu (k_i,k_{{\rm ref},i}) & \equiv \frac{\langle k_{{\rm ref},i}| \gamma_\mu | k_i ]}{\sqrt{2}\langle k_{{\rm ref},i} k_i  \rangle }\\ 
 (\zeta_i^-)_\mu (k_i,k_{{\rm ref},i}) & \equiv \frac{- [ k_{{\rm ref},i}| \gamma_\mu | k_i \rangle}
 {\sqrt{2}[ k_{{\rm ref},i} k_i  ] }\,.
\end{split}\end{equation}
We have here employed 
\begin{equation}
| k_j \rangle \equiv u_+(k_j) \equiv \Big[\frac{1}{2}(1+\gamma_5)\Big]u(k_j)\,,
\end{equation}
\begin{equation}
\langle k_j | \equiv \bar u_-(k_j)\equiv u(k_j)\Big[ \frac{1}{2}(1+\gamma_5)\Big]\,,
\end{equation}
\begin{equation}
| k_j ]\equiv u_-(k_j)\equiv \Big[\frac{1}{2}(1-\gamma_5)\Big]u(k_j)\,,
\end{equation} 
\begin{equation}
[ k_j | \equiv \bar u_-(k_j) \equiv \bar u(k_j)\Big[\frac{1}{2}(1-\gamma_5)\Big]\,,
\end{equation}
furnished with spinor solutions to Dirac's equation represented by $u(k_j)$ and $\bar u(k_j)$. We remark that the reference momentum in the overhead expressions can be taken arbitrarily. Specifying the spinor products 
\begin{equation}
\bar u_- (k_i) u_+ (k_j) = \langle ij\rangle,\quad \bar u_+ (k_i) u_- (k_j) = [ ij],
\end{equation}
we have realized that in gauge theories one can magically express $n$-point amplitudes in one simple formula dubbed the Parke-Taylor amplitude.
\begin{equation}
A_n^{\rm tree}(1^-,2^-,\ldots,n^+)= i \frac{\langle 1 2 \rangle^4}{\langle 1 2\rangle \langle 2 3\rangle\ldots \langle n 1 \rangle}.
\end{equation}
This formalism can be utilized for gravity by employing the concept of building gravity amplitudes using the KLT technique. It has its roots in the manner string theory amplitudes are organized perturbatively, and has been pioneered by Bern et al. in numerous gravity contexts \cite{Bern:2002kj}.\\[5pt] 
The KLT relations for gravity amplitudes take the following generic form for $n$-point scattering using a compact formulation in terms of the $S$-kernel~\cite{Bjerrum-Bohr:2010diw,Bjerrum-Bohr:2010pnr}. Here  $A^{\rm
  tree}(n,n-1,\sigma(2,\cdots,n-2),1)$  signifies gauge theory amplitudes. %
\begin{multline}
  iM^{\rm tree} =\sum_{\tau, \gamma\in  S_{n-3}} \mathcal S[\tau(2,\cdots,n-2)|\tau(2,\cdots,n-2)]|_{k_1}\times
\cr
\label{e:KLT}\times  A_n^{\rm tree}(1,\tau(2,\cdots,n-2),n-1,n) A_n^{\rm
    tree}(n,n-1,\sigma(2,\cdots,n-2),1)  \,.
\end{multline}
with $S_{n-3}$ representing the possible permutations over $n-3$ indices and where
 the momentum kernel $\mathcal S$ is supplied by the expression ($\theta(i_t,i_s)$ equals 1 if the ordering of the legs $i_r$ and $i_s$
is opposite in the sets $\{i_1,\dots,i_r\}$ and $\{j_1,\dots,j_r\}$, and 0 if the
ordering is the same.)
\begin{equation}
  \label{e:Skernel}
 \mathcal S[i_1,\dots, i_r|j_1,\dots, j_r]|_p=\prod_{t=1}^r (p\cdot
  k_{i_r}+\sum_{s>t}^r \theta(i_r,i_s)\, k_{i_r}\cdot k_{i_s} )\,.
\end{equation}
Thus utilizing this formula $n$-point gravity amplitudes
$M_n^{\rm tree}$ can be delivered from color-ordered amplitudes in gauge theory
$A_n^{\rm tree}(\sigma)$. The above form of the KLT relations is frequently favored in unitarity computations but it is critical to recognize that multiple identical rewritings of the KLT relations are feasible. This is due to the equivalences that follow from
\begin{equation}
\sum_{n\in S_{n-2}} \, \mathcal  S[\tau(2,\cdots,n-1)|\sigma(2,\cdots,n-2)]|_{k_1}
\times  A_n^{\rm tree}(1,\tau(2,\cdots,n-1),n)=0; \
\forall\sigma\in S_{n-2}\,,
\end{equation} 
Thus with the  KLT approach, we can efficiently evaluate the emission of two gluons from massive scalars without having to consider graviton vertex rules. From these relations, it is possible to derive helicity expressions for graviton amplitudes which can be employed as input in generalized unitarity cuts.

Exploiting the spinor helicity formalism we generate very compact identities for graviton amplitudes. Graviton helicity polarisations can be inferred from gluon polarisations as follows
\begin{equation}\begin{split}
(\xi_i^+)_{\mu\nu} (k_i,k^L_{{\rm ref},i},k^R_{{\rm ref},i}) & = (\zeta_i^+)_\mu (k_i,k^L_{{\rm ref},i}) \times (\zeta_i^+)_\nu (k_i,k^R_{{\rm ref},i})\\ 
 (\xi_i^-)_{\mu\nu}  (k_i,k^L_{{\rm ref},i},k^R_{{\rm ref},i}) & =  (\zeta_i^-)_\mu (k_i,k^L_{{\rm ref},i})\times (\zeta_i^-)_\nu (k_i,k^R_{{\rm ref},i})\end{split}\end{equation}

Employing these polarisation vectors for computations of the two-scalar-two-graviton amplitude we attain the compact expressions
\begin{eqnarray}
iM_{0}^{\rm tree}(p_1,p_1',k_1^+,k_2^+) &=&{\kappa_{(4)}^2\over 16}\,
{1\over (k_1\cdot k_2) }\,
{m^4 \sq[k_1,k_2]^4\over (k_1\cdot p_1)(k_1\cdot p_1')} \,, \cr
iM_{0}^{\rm tree}(p_1,p_1',k_1^-,k_2^+)&=&{\kappa_{(4)}^2\over16}\,{1\over (k_1\cdot k_2)}\,
{ \spab[k_1,p_1,k_2]^2\spab[k_1,p_1',k_2]^2\over (k_1\cdot p_1)(k_1\cdot p_1')}\,,
\end{eqnarray}
with 
$ iM_{0} ^{\rm tree}(p_1,p_2,k_1^-,k_2^-)=(iM_{0} ^{\rm
tree}(p_1,p_2,k_1^+,k_2^+))^*\,,
$
and 
$iM_{0}^{\rm tree}(p_1,p_2,k_1^+,k_2^-)=(iM_{0} ^{\rm
tree}(p_1,p_2,k_1^-,k_2^+))^*.
$\\[5pt]
For the two-scalar-three-graviton amplitude $(+++)$ configuration we see that
\begin{equation}\begin{split}
i\mathcal M_0^{\rm tree}&(p_1,p_1',k_1^+,k_2^+,k_3^+)  =\cr & \bigg(-\frac{(8\pi G_N)^{3\over2} m_1^4}{\an[k_1,k_2]^2 \an[k_1,k_3]^2 \an[k_2,k_3]^2} \sum_{1 \leq i \ne j \ne k \leq 3} {(k_i\cdot k_j)(k_j\cdot k_k) tr_{+}[k_k,p_1,p_1',k_i]\over
    (p_1\cdot k_k)(p_1'\cdot k_i)}\bigg),
\end{split}\end{equation}
with $i\mathcal M_0^{\rm tree}(p_1,p_1',k_1^-,k_2^-,k_3^-)$ obtained by complex conjugation. This expression agrees with~\cite{Bern:1998sv} and vanishes as expected when $m_1=0$. We have defined 
\begin{equation}
  tr_\pm(abcd)\equiv 2(a\cdot b c\cdot d-a\cdot c b\cdot d+a\cdot d b\cdot
  c)\pm 2i \epsilon^{\mu\nu\rho\sigma}a_\mu b_\nu c_\rho d_\sigma\,.
\end{equation}
For the $(-++)$ amplitude we get
\begin{equation}\begin{split}
   \label{e:tggGGGhelmpp}
i\mathcal M_0^{\rm tree}&(p_1,p_1',k_1^-,k_2^+,k_3^+)  \cr & ={ (2\pi G_N)^{3\over2}\over2}
 \Bigg(\sum_{2 \leq j \ne k \leq 3} \frac{
   \spab[k_1,p_1,k_j]\spab[k_1,p_1',k_j]^2\spab[k_1,p_1,k_k]^3}{\an[k_1,k_j]\an[k_1,k_k](k_1\cdot
   k_j)(k_1\cdot k_k)(p_1\cdot k_1)(p_1'\cdot k_j)} \cr &
 - \frac{
   \spab[k_1,p_1,k_2]^3\spab[k_1,p_1',k_3]^3}{\an[k_1,k_2]\an[k_1,k_3](k_1\cdot
   k_2)(k_1\cdot k_3)(p_1\cdot k_2)(p_1'\cdot k_3)}\cr &- \frac{2
   \sq[k_2,k_3] \spab[k_1,p_1,k_2]\spab[k_1,p_1,k_3]\langle k_1 | p_1
   | p_1' | k_1\rangle ^2}{\an[k_1,k_2]\an[k_1,k_3]\an[k_2,k_3](k_1\cdot k_2)(k_1\cdot k_3)(p_1\cdot k_1)}
 \cr &+\frac{2 \sq[k_2,k_3]^3 \langle k_1 | p_1 | p_1' | k_1\rangle^2}{\an[k_2,k_3](k_1\cdot k_2)(k_1\cdot k_3)t} \Bigg)+(p_1 \leftrightarrow -p_1'),
\end{split}\end{equation}
with $i\mathcal M_0(p_1,p_1',k_1^+,k_2^-,k_3^-)$ obtained by complex conjugation.\\[5pt]
We will assess how these tree amplitudes are useful when we employ unitarity cuts as a route to loop amplitudes. We comment in passing that the above amplitude examples would be extremely cumbersome to derive straight from the vertex rules. This demonstrates the power of the double-copy KLT techniques fused with spinor-helicity. An interesting perspective arises when considering the Kawai-Lewellen-Tye (KLT) relations in the context of effective field theory extensions. In many cases, it is possible to maintain KLT factorizations of higher derivative terms in generic effective Lagrangians for gauge theories and gravity. This opens up simpler amplitude computations of extensions of Einstein's gravity. More details on such calculations can be found in refs. \cite{Bjerrum-Bohr:2003utn,Bjerrum-Bohr:2003hzh,Bjerrum-Bohr:2004vlu,Bjerrum-Bohr:2010eeh}.

\section{New on-shell methods and double-copy and BCJ numerators}
Current techniques have further simplified the framework for the computation of tree amplitudes of gravitational interactions. We will in this section view how an even better compact construction for graviton amplitudes can be enabled from color-kinematics numerators. Utilizing such numerators it has been established that classical results in pure gravity can be derived in a dimension agnostic way. A path to this framework is from the computation of tree-level multi-graviton emission from a scalar line by utilizing the scattering equations procedure pioneered by refs.
\cite{Cachazo:2013hca,Cachazo:2013iea,Cachazo:2013gna,Cachazo:2014nsa}. This formalism reaches scattering amplitudes automatically satisfying color-kinematic dualities for a large class of field theories in a remarkably compact manner. Amplitudes for gluon amplitude acquired employing this framework are developed from contour integrals. See, ~\cite{Bjerrum-Bohr:2014qwa} for additional details.
%
\\[5pt]
The important ingredient in such formulations is production of numerator\\ $N_{n-2}(1,\beta(2,\dots,n-1),n)$ which can be written in terms of tensor products of gluon polarisation vectors and momenta. Several efficient schemes for computing such numerators have been developed, see {\it e.g.}~\cite{Fu:2017uzt,Teng:2017tbo,Bjerrum-Bohr:2019nws,Bjerrum-Bohr:2020syg}.  %
Using this one arrive at scattering amplitude 
 \begin{equation}\begin{split}
& A^{\rm tree}_{n-2}(1,\sigma(2,\dots,n-1),n)=\  \sum_{\gamma\in S_{n-2}}\!\!\!\!
\mathcal S^{-1}(\beta\gamma)|_{p_1}N_{n-2}(1,\sigma(2,\cdots,n-1),n)\,,  
\end{split} \end{equation}
where $S^{-1}(\beta\gamma)|_{p_1}$ denotes the associated denominator poles. This function can be understood as the inverse momentum kernel. Using this in the context of gravity using the KLT relations one arrives at
 \begin{equation}\label{e:MtreeKLTCHY}
{M}^{\rm tree}_{n-2}(1,2,\ldots,n) = 
i\,\sum_{\sigma\in \ S_ {n-2}} N_{n-2}(1,\sigma(2,\cdots,n-1),n) {
 A}_{n-2}(1,\sigma(2,\dots,n-1),n) \,,
\end{equation} 
which is a compact starting point for the computation of tree graviton amplitudes. It is essential for the efficient usage of these formulas that poles are solely introduced into the expression from the Yang-Mills amplitude on the right.
From dimensional reduction one can derive multi-gluon emission from a massive scalar from pure multi-gluon amplitudes. Thus this formalism is remarkably malleable and one can use it in the context of several different theories, for instance, with massive spin-1 or fermion,  states. See \cite{Bjerrum-Bohr:2020syg} for some additional details. %
%
%
Some numerator and amplitude examples are 
\begin{equation} \label{e:M3pt}
M^{\rm tree}_{1}(p,\ell_2,-p')=i\,N_1(p,\ell_2, -p')A_1(p,\ell_2, -p')=i\,N_1(p,\ell_2, -p')^2 ,
\end{equation}
as well as 
\begin{equation} \label{e:M4pt}\begin{split}
M^{\rm tree}_{2}(p,\ell_2,\ell_3,-p')
= i\,N_{2}(p,2,3,-p') {A}_{2}(p,2,3,-p')+ {\rm perm.} \{2,3\}\\ 
=\frac{{i N_2}(p,2,3,-p')^2}{(\ell_2+p)^2-m^2+i\varepsilon}+ \frac{{i N_2}(p,3,2,-p')^2}{(\ell_3+p)^2-m^2+i\varepsilon}+\frac{i({N_2}^{[2,3]})^2}{(\ell_2+\ell_3)^2+i\varepsilon}\,,
\end{split}\end{equation}
and we have comparable expressions at higher multiplicities. We direct
to this online
\href{https://nbviewer.org/github/pierrevanhove/Probe/blob/main/probe.ipynb}{repository}
for expressions for numerator factors. We emphasize that amplitudes derived this way are accurate in arbitrary dimensions, satisfy manifestly color-kinematic relations, and incorporate no spurious poles. We can equally well generate expressions for multi-photon emission from a massive charged scalar exploiting the photon decoupling identity
\begin{multline}
 A^{\rm photon}_{n-2}(p,2,\dots,n-1,-p')=\sum_{\beta\in S_{n-2}} A^{\rm tree}_{n-2}(p,\beta(2,\dots,n-1),-p')\,.\hfill
 \end{multline}
This can be utilized for computations in scalar QED. 
\subsection{Remodelling gravitational amplitudes}
The emphasis here is on the application of modern amplitudes
methods for evaluating the gravitational interaction between two
bodies and their association to gravitational physics.\\[5pt]
Current advancement has materialized from utilizing generalized unitarity. In such studies, we require 1) to acquire an ansatz for the considered amplitude in the context of a basis of master integrals, and 2) coefficients for each master integral assessed by considering unitarity conditions. We have in the earlier section witnessed how to compute input trees for unitarity efficiently.
The recent
improvement in computing on-shell amplitudes permits  evaluating loop amplitudes at high perturbative
orders thereby acquiring relativistic invariant results valid in all
energy regimes from small relative velocities to the ultra-relativistic limit~\cite{Damour:2017zjx,DiVecchia:2020ymx,Bjerrum-Bohr:2021vuf}. Deriving the classical post-Minkowskian
results as a component of the full quantum gravity $S$-matrix between two
massive bodies open a new stance on 
the subtle queries for instance on gravitational radiation and ultra-high
energy scattering in classical gravity.
The scattering amplitude method completes the post-Newtonian
computations by delivering information exceeding its regime of validity
and leads to astonishing
results joining the conservative part and
gravitational radiation effects~\cite{Damour:2017zjx,Damour:2019lcq,Damour:2020tta,Parra-Martinez:2020dzs,DiVecchia:2021ndb,DiVecchia:2021bdo,Herrmann:2021tct,Bjerrum-Bohr:2021vuf}. It offers a new outlook on the traditional
methods~\cite{Goldberger:2007hy,Goldberger:2022ebt,Goldberger:2022rqf,Blanchet:2013haa,Porto:2016pyg,Barack:2018yly,Isoyama:2020lls,Buonanno:2022pgc}
 employed to compute the gravitational-wave templates.
The same formalism unites the re-summed
 post-Newtonian results~\cite{Bern:2019nnu,Bern:2019crd,Bern:2021dqo,Bern:2021yeh}
and the high-energy
behaviour~\cite{DiVecchia:2020ymx,DiVecchia:2021bdo}.\\[5pt]
 Another attractive characteristic is that the scattering amplitude approach
 can be applied to any space-time dimension. This is especially
 intriguing since black holes in higher dimensions offer
 fascinating new features~\cite{Emparan:2008eg}.\\[5pt]
The extraction of the
classical General Relativity contribution from scattering amplitudes
has guided a reformulation of the classical $\hslash\to0$ limit~\cite{Bjerrum-Bohr:2021vuf} of
scattering amplitudes that differ from the logic of the na\"ive textbook $\hslash$ counting
e.g.~\cite{Itzykson:1980rh}.  
This approach is an excellent exposition of the
 statement by Kovacs and Thorne 
in~\cite{Kovacs:1978eu} that \emph{``Any classical problem can be solved
quantum-mechanically; and sometimes the quantum solution is easier
than the classical.''}
\subsection{Setting up a convenient formalism}
We will now consider establishing a convenient convention for computations. We are interested in extracting physical observables from the
gravitational interactions between two massive body of masses $m_i$ and spin $S_i$ with $i=1,2$ interacting through the exchange of massless spin-2 gravitons~\cite{tHooft:1974toh,Veltman:1975vx,DeWitt:1967yk,DeWitt:1967ub,DeWitt:1967uc}.\footnote{One could
as well include electro-magnetic interactions as considered
in~\cite{Donoghue:2001qc,Bjerrum-Bohr:2002aqa}. The vital limitation of the present
analysis is that we have large external masses or charges with a
massless exchange. Observables in scalar QED have been regarded as a
testing ground for the gravitational interactions~\cite{Bern:2021xze}.} 
The two-body scattering matrix can be expanded in a perturbation series
\begin{equation}
{\mathcal M}(p_1,p_2,p_1',p_2')=\!\!\!\!\begin{gathered}
    \includegraphics[width=2.5cm]{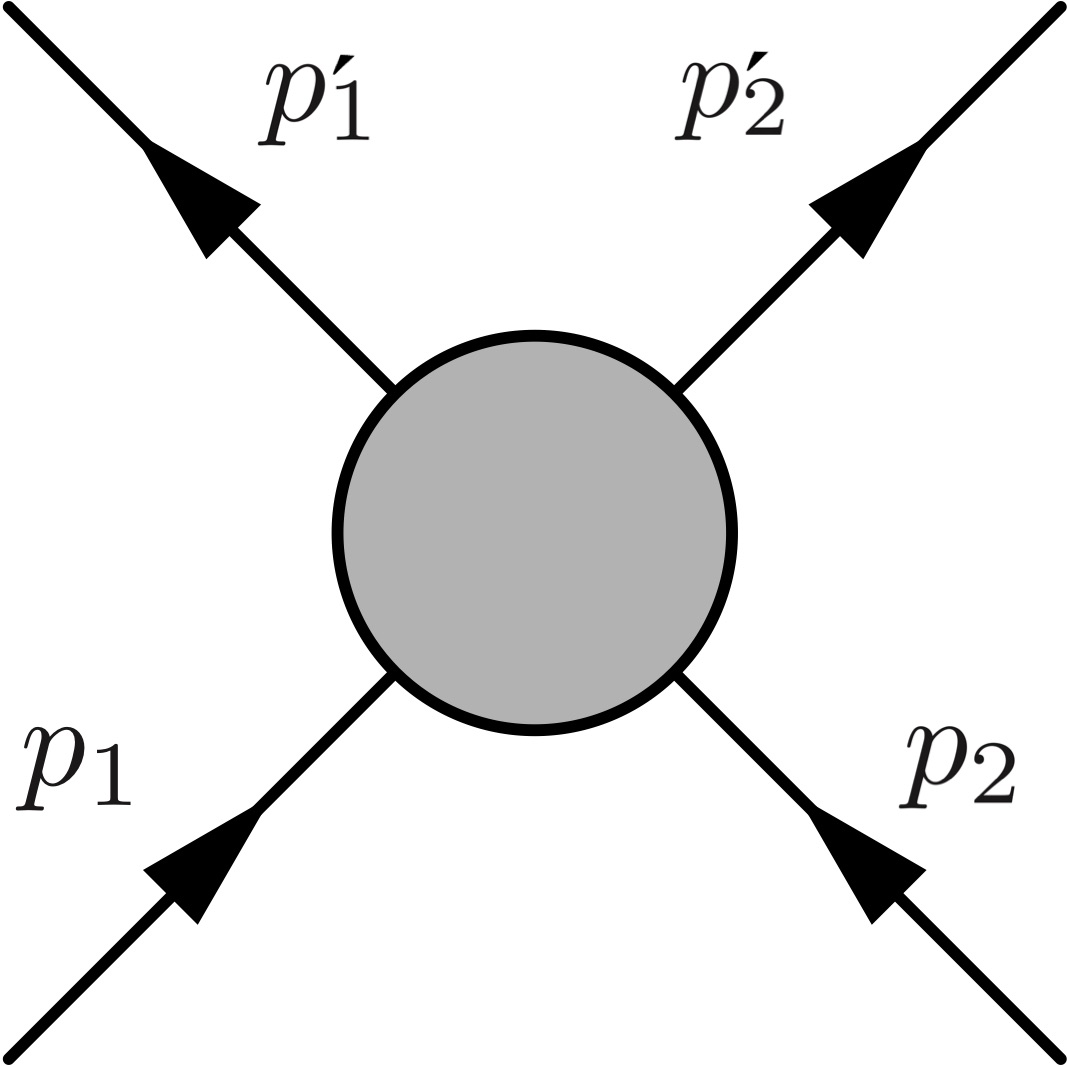}
  \end{gathered}
\!\!\!\!=\sum_{L\,=\,0}^{ \infty} G_N^{L+1} \mathcal{M}_{ L}(\gamma, q^2) .
\end{equation}
The quantum scattering matrix ${\mathcal M}(p_1,p_2,p_1',p_2')$
depends on the relativistic  factor $\gamma:=p_1\cdot p_2/(m_1m_2)$, the
momentum transfer $(p_1-p_1')^2=:q^2$ and $\hslash$.\\[5pt]
At a given order in perturbation, one obtains the exchange of gravitons
(curly lines) between massive external matters (solid lines)
$$\includegraphics[width=11cm]{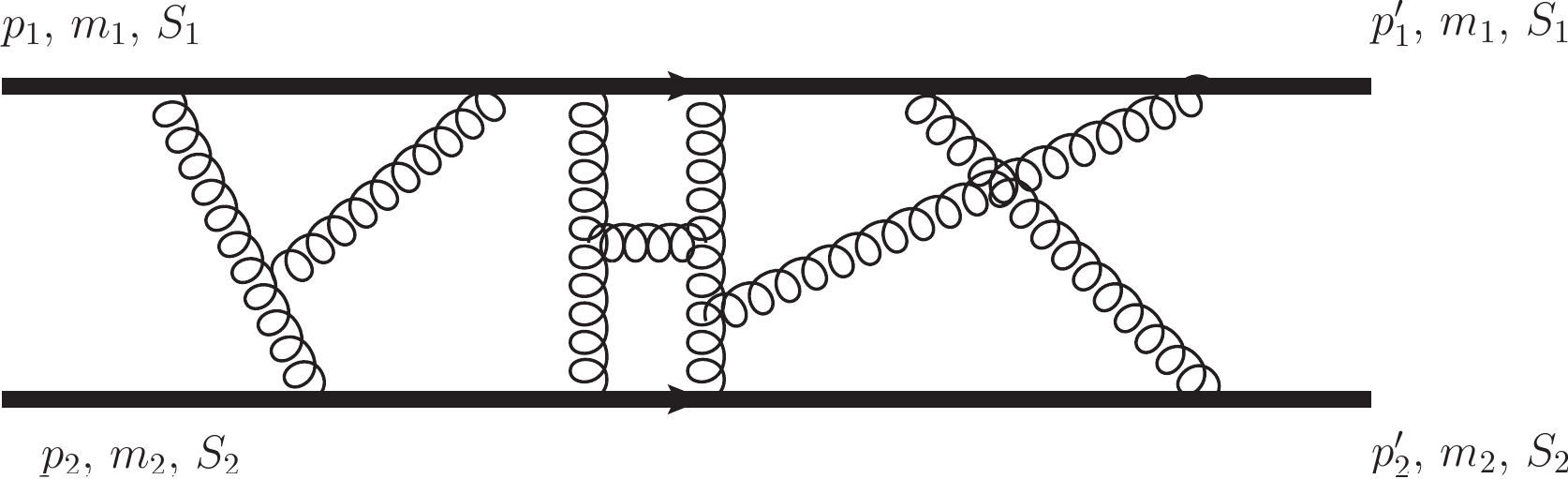}$$
\subsubsection{The examples of tree-level and one-loop amplitudes}
We illustrate the emergence of the classical and quantum pieces at
tree-level and one-loop order.
\begin{figure}[ht]
  \centering
   \includegraphics[width=5cm]{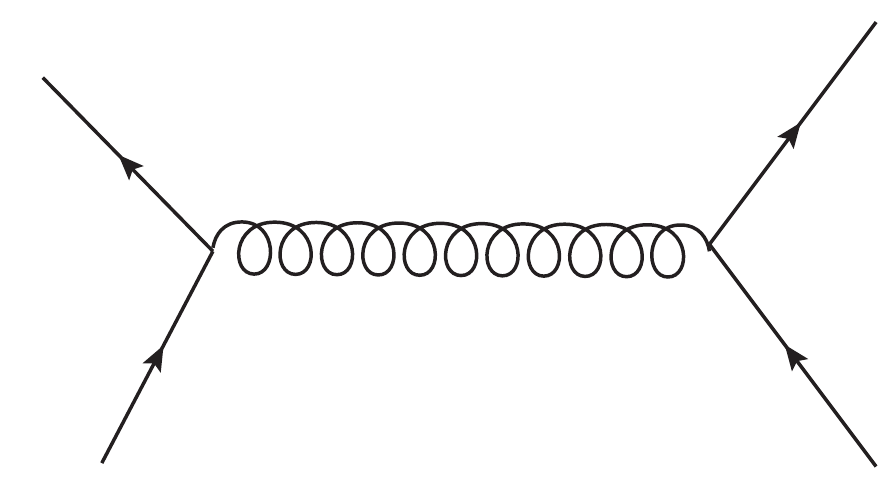}
  \caption{Gravitational two-body scattering at tree-level}
  \label{fig:tree}
\end{figure}
The tree-level scattering between two massive fields in
fig.~\ref{fig:tree} has the $\hslash$ expansion 
\begin{equation}\label{e:M0}
\mathcal M_0(\gamma,\underline q^2,\hslash)= {-16\pi G_N m_1^2m_2^2 \left(2\gamma^2-1\right)\over {  \hslash} |\underline q|^2}+{  \hslash}{     4\pi G_N p_1\cdot p_2}.
\end{equation}
In this expression, we recognize the first Post-Minkowskian
classical contribution $G_N m_1^2m_2^2(2\gamma^2-1)$ matching the
results from General Relativity see e.g.~\cite{Damour:2019lcq}, and
the higher order quantum correction $\hslash p_1\cdot p_2$ due to the
contact term as mandated by the full quantum amplitude.\\[5pt]
In ref. \cite{Bjerrum-Bohr:2013bxa} we elucidated that generalized unitarity is an excellent tool to calculate terms that resemble long-range contributions in the amplitude. Such non-analytic terms furnish us with classical scattering potentials in theories such as QED and gravity along with quantum modifications. Since we are exclusively interested in non-polynomial contributions it is not demanded that we generate the full amplitude, as an identification of those terms in the amplitude that is non-analytic is adequate for classical and leading quantum corrections. Thus a pathway is established to streamline such computations. At one-loop order we fetch coefficients corresponding to $1/\sqrt{-q^2}$ and $\log(-q^2)$ terms in the amplitude from on-shell unitarity. Proceeding following the approach provisioned in~\cite{Bern:1994cg}, this can {\it e.g.} be done through evaluating the phase-space integrals by reinstating the off-shell cut propagators with on-shell cut conditions in numerators.\\[5pt]
\begin{figure}[h]
\centering\includegraphics[width=6cm]{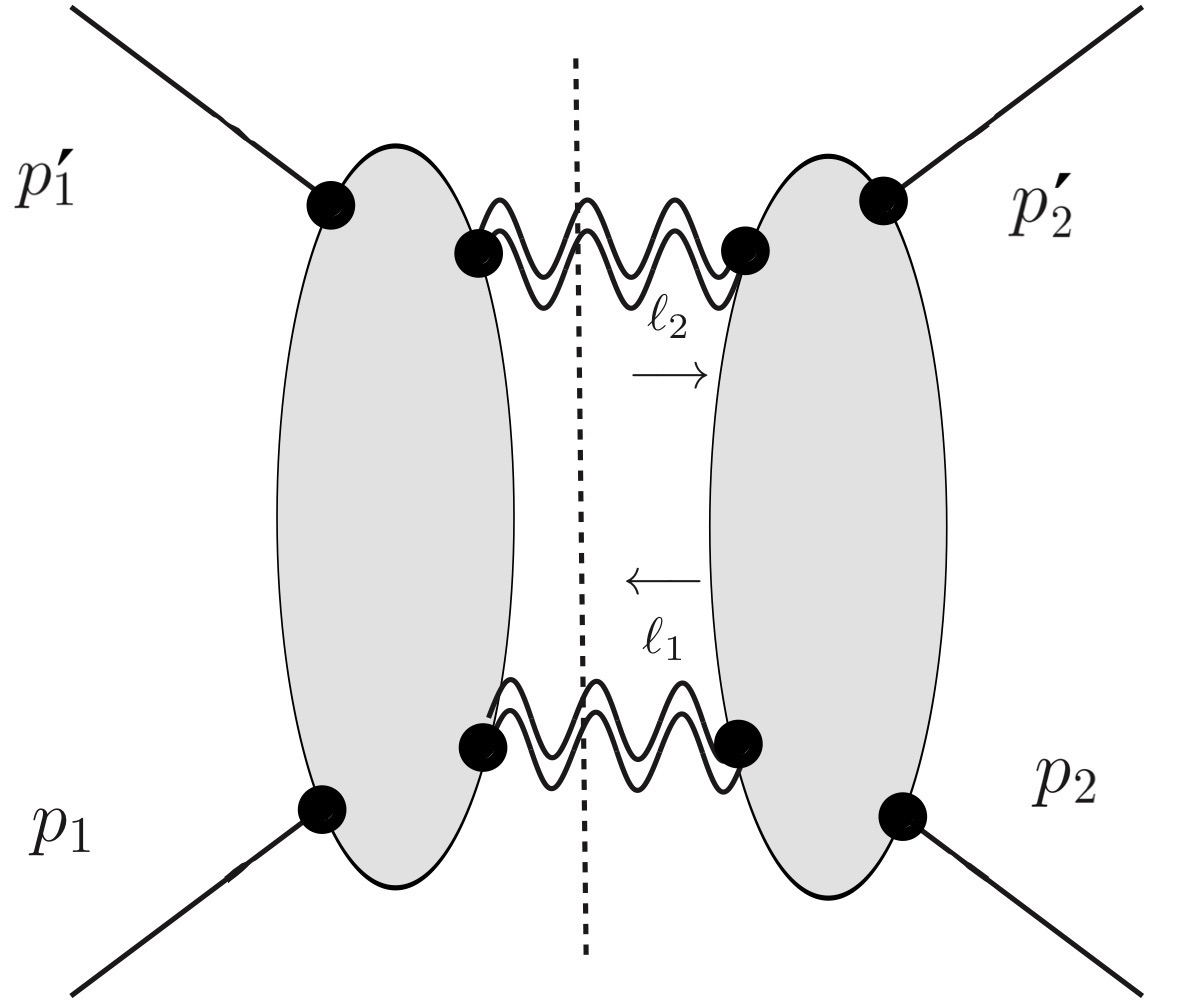}
\caption{The double cut.}\label{fig:cut}
\end{figure}

Formally at one loop, we thus have to consider the cut in fig.~\ref{fig:cut}
associated with the integral expression
  \begin{equation}
  \label{e:dischel}
\left.  iM^{\rm 1-loop}\right|_{disc}=\int {d^D\ell\over (2\pi)^D}\,
 {\sum_{\lambda_1,\lambda_2} M^{\rm
    tree}_{\lambda_1\lambda_2}(p_1,p_1',-\ell_2^{\lambda_2},\ell_1^{\lambda_1})
 (M^{\rm
    tree}_{\lambda_1\lambda_2}(p_2,p_2',\ell_2^{\lambda_2},-\ell_1^{\lambda_1}))^* \over \ell_1^2 \ell_2^2}\Big|_{cut}
  \,,
\end{equation}
Here we have the cut conditions $\ell_1^2=\ell_2^2=0$ and we sum over all feasible physical graviton helicity arrangements across the cut: $\lambda_1$ and $\lambda_2$. At one loop, the basis for the quantum and classical contributions is supplied by box, triangle, and bubble graph topologies. In the cut, we can directly pinpoint the integral functions and thus isolate the coefficients for the non-analytic terms that we are pursuing. In \cite{Bjerrum-Bohr:2021din} we isolated the classical non-analytically contributions by evaluating the triple cut and recognizing the coefficients for the two-loop basis of integral functions.
At the one-loop order the full
quantum two-body scattering amplitude is thus expanded on the standard
basis of scalar one-loop integrals in four dimensions
\begin{equation}\label{e:M1}
  \mathcal{M}_1=\frac{i16 \pi^2
    G_N^2 m_1^2m_2^2}{E_1E_2}\bigg(
  3 \big(1-5\gamma^2\big) \left(
  m_1^2  {I}_{\triangleright}+ m_2^2 {I}_{\triangleleft}\right)
+4m_1^2m_2^2 \big(1-2\gamma^2\big)^2
  \left( {I}_{\Box}+ {I}_{ \bowtie}\right) +\cdots \bigg),
\end{equation}
where $E_i^2=\vec p_i^2+m_i^2$ with $i=1,2$ is the energy of the
particle $i=1,2$.
The expression contains the scalar  massive triangles which have the
large mass expansion exhibiting the classical $1/\hslash$ term~\cite{Iwasaki:1971vb,Holstein:2004dn,Bjerrum-Bohr:2018xdl}
 \begin{equation}
 {I}_{\triangleright}=-\frac{i}{32
                          m_1 }\frac{1}{\hslash|\underline{q}\,|}+\cdots ,\qquad
                          {I}_{\triangleleft}=-\frac{i}{32
  m_2}\frac{1}{\hslash|\underline{q}\,|}+\cdots
 \end{equation}
and the scalar box and cross-box integrals
 \begin{align}
{I}_{\Box}
                    &=\frac{i}{16\pi^2\hslash^2|\underline{q}\,|^2}\left(-\frac{1}{m_1m_2}+{m_1(m_1-m_2)\over
                      3m_1^2m_2^2}+{   {i\pi \over |p|(E_1+E_2)}}\right)\bigg({  \frac{2}{3-d} }-\hslash^2\log|q|^2\bigg)+\cdots\cr
{I}_{\bowtie}
                    & =\frac{i}{16\pi^2\hslash^2|\underline{q}\,|^2}\left(\frac{1}{m_1m_2}-{m_1(m_1-m_2)\over
                      3m_1^2m_2^2}\right)\bigg({  \frac{2}{3-d} }-\hslash^2\log|q|^2\bigg)+\cdots
 \end{align}
putting everything together we get that expansion of the full quantum
one-loop amplitude to read
 \begin{multline}
\mathcal{M}_1(\gamma,\underline q^2,\hslash)=\frac{\pi^2
  G_N^2m_1^2m_2^2}{E_1E_2}\bigg[-\frac{3 \big(1-5\gamma ^2\big)}{2\hslash|\underline{q}\,|} (m_1+m_2)\cr
+{  
\frac{im_1m_2}{(E_1+E_2)}\frac{4
  \big(1-2\gamma^2\big)^2}{|\vec{p}\,|}\frac{(\frac{2}{3-d}-\hslash^2\log|{q}|^2)}{\pi\hslash^2|\underline
  q|^2}}\bigg]+\cdots\,.
\end{multline}
This expression contains
\begin{itemize}
  \item at order $1/\hslash^2$ a contribution given by the square of
    the classical tree-level contribution from~\eqref{e:M0}. This
    piece is needed for the exponentiation of the $S$-matrix
    in~\eqref{e:NtoT} as detailed in~\cite{Damgaard:2021ipf}. 
\item At order $1/\hslash$ the classical second
post-Minkowskian contribution~\cite{Bjerrum-Bohr:2018xdl} matches the classical second
post-Minkowskian result for generic masses.
\item A quantum piece of order $\hslash^0$ which is a long-range
  infrared quantum gravity effect. Because this is the first quantum
  correction to the classical result the value of the quantum gravity-induced correction is universal and independent of the ultraviolet
  regularisation~\cite{Bjerrum-Bohr:2013bxa,Bjerrum-Bohr:2014zsa}.
\end{itemize}  
We will now generalize this discussion to any perturbative order. A traditional argument (see for instance~\cite{Itzykson:1980rh}) gives that the $L$-loop  contribution is of order 
$ \mathcal M_L(\gamma,q^2)=\mathcal O(\hslash^{L-1} )$. 
A different behavior emerges when maintaining fixed
wave-number $\underline q= q/\hslash$ and taking both the
$\hslash\to0$ and the small momentum transfer $q\to0$
limit~\cite{Holstein:2004dn,Bjerrum-Bohr:2018xdl,Kosower:2018adc}.
The $L$-loop two-body scattering amplitude has the Laurent expansion
in four dimensions
\begin{equation}\label{e:Mhslashexp}
     \mathcal M_L(\gamma,\underline q,\hslash)={\mathcal
       M_L^{(-L-1)}(\gamma,D)\over
       \hslash^{L+1}|\underline q|^{{L(4-D)\over2}+2}}+\cdots +{\mathcal
       M^{(-1)}_L(\gamma,D)\over \hslash |\underline q|^{{L(4-D)\over2}+2-L}}+O(\hslash^0).
   \end{equation}
   Again, we can systematize the full quantum amplitude in terms of three types of contributions:
   \begin{enumerate}
   \item  the
terms of order $1/\hslash^r$ with $L+2\leq r\leq 2$ that are more
singular than the classical piece in the $\hslash\to0$ limit. 
\item the classical piece of order
$1/\hslash$ from which the classical Einstein gravity contribution is
extracted. The presence of a classical piece in the quantum
gravitational two-body amplitude was first shown at one-loop
by~\cite{Iwasaki:1971vb}, and this is articulated as an all order
statement in~\cite{Holstein:2004dn}. 
The expansion in~\eqref{e:Mhslashexp} is unusual but this is a natural
when considering a large external mass expansion of the two-body
gravitational scattering. 
At the $L+1$ post-Minkowskian order, the  two-body scattering amplitude between two
massive particles have the following mass dependence
\begin{equation}\label{e:MLclassical}
\mathcal M_L(\gamma,q^2)= {{     G_N^{L+1} }m_1^2m_2^2\over
 \underline q^{2+{(2-D)L\over2}}}\sum_{i=0}^L c_{L-i+2,i+2}(\gamma) {    
  m_1^{L-i} m_2^i}.
\end{equation}

This classical contribution  emerges from the $1/\hslash$ piece of the
quantum amplitude in~\eqref{e:Mhslashexp} is one remembers that the
mass dependence in quantum field theory  appears as the Compton
wave-length $mc/\hslash$. Expressing the classical contribution by
making this explicitly gives
\begin{equation}
\mathcal M_{L}(\gamma,q^2,\hslash)=\cdots+ \underbrace{ {m_1^2m_2^2\over
  \underline q^{2+{(2-D)L\over2}}}\,  {  \hslash}^{L-1}
 \, {     G_N^{L+1} }\sum_{i} \left({     m_1}c\over {  \hslash}
     \right)^{L-i} \left( {     m_2}c\over {  \hslash}
      \right)^{i}}_{={\mathcal M_L(\gamma,\underline q^2)\over {  \hslash}} }+\cdots 
  \end{equation}
Therefore the polynomial
mass dependence of the classical amplitude in~\eqref{e:MLclassical}
 expected 
for the conservative part of the scattering
angle~\cite{Antonelli:2019ytb,Bini:2020hmy} arises consistently from
the classical limit of the quantum amplitude.
  The $\underline q^2$ dependence of the classical contributions are
exactly what one anticipates to contribute to the three dimensional
potential at the $L+1$ post-Newtonian order since
\begin{equation}
 G_N^{L+1} \int d^3\vec{\underline q}  {e^{i\underline q\cdot \vec r}\over
  |\underline q|^{2-L}}\propto \left(G_N\over r\right)^{L+1}.
\end{equation}
  The expansion
in~\eqref{e:Mhslashexp} indicates that a given $\hslash$ order has a prescribed
analytic dependence in $q^2$. This was utilized to elucidate that the
classical contribution can be identified from specific unitarity
cuts~\cite{Bjerrum-Bohr:2018xdl}.    The extraction of the classical part has
been since systematized using  an
heavy-mass effective theory
approach~\cite{Brandhuber:2021kpo,Brandhuber:2021eyq}, or the velocity cut
formalism~\cite{Bjerrum-Bohr:2021vuf,Bjerrum-Bohr:2021din,Bjerrum-Bohr:2021wwt} which is
discussed further in section~\ref{sec:newmethods}.  
\item the quantum corrections of order $\hslash^r$ with
$r\geq0$, which leads to quantum gravity corrections to the classical
Einstein gravity results~\cite{Bjerrum-Bohr:2014zsa}.
\end{enumerate}
\subsection{Scattering potentials from the Lippmann Schwinger equation and gravitational eikonalisation}
Supplied with the scattering amplitude we can relate to a Hamiltonian scattering potential via numerous avenues, for instance, one can derive a classical Hamiltonian by employing Born subtractions in the framework of the Lippmann-Schwinger equation~\cite{Cristofoli:2019neg, Kalin:2019rwq,Bjerrum-Bohr:2019kec}. 
In this setup, one links the relativistic scattering amplitude $\mathcal M$ to the potential $\mathcal V$, through: 
\begin{equation}
\tilde{\mathcal{M}}_L(p,p') ~\equiv~ \frac{\mathcal{M}_L(p,p')}{4E_1E_2}\,,
\end{equation}
with 
\begin{equation}
p_1=(E_1,\vec p),\quad p_1'=(E_1, \vec p\,'),\quad p_2=(E_2, -\vec p), \quad p_2'=(E_2, -\vec p\,')\,, 
\end{equation}
\begin{equation}
    \mathcal V(r,p)= \int {d^3q\over(2\pi)^3} e^{i q\cdot r} \mathcal V(p,q) = \int {d^3q\over(2\pi)^3} e^{i q\cdot r} \tilde{\mathcal{M}}_0(p,q)\,.
\end{equation}
\begin{equation}
  \tilde{\mathcal M}(p,p') = \mathcal V(  p,p')+ \int
  {d^3k\over (2\pi)^3}   { \mathcal V(
    p,k) \mathcal M(k,p')\over E_p-E_k+i\varepsilon}\,, \label{Lippmann}
\end{equation}
This extension, through a one-particle
Salpeter equation explains simply and systematically -- the subtractions mandated -- in principle to any perturbative order.
However, at high perturbative loop order, it is an arduous task to keep track of the multiple new computational pieces concerned (at this point in writing exclusively two-loop order has been carefully examined). An alternative framework for determining a Hamiltonian is using the effective field theory matching strategy of ~\cite{Cheung:2018wkq,Bern:2019nnu,Bern:2019crd,Cheung:2020gyp,Bern:2021dqo,Bern:2021yeh}. The parallelism between these two systems can be ascertained to hold on general grounds~\cite{Cristofoli:2020uzm}.

Recourse to a Hamiltonian framework for computing observables is the eikonal formalism.
Employing the eikonal formalism one converts the amplitude following
\begin{equation}\label{e:FT}
  \mathcal M_L(\gamma,b,\hslash)={1\over 4m_1m_2\sqrt{\gamma^2-1}} \int_{\mathbb
    R^{D-2}} {d^{D-2}\vec {\underline q}\over (2\pi)^{D-2}} \mathcal
  M_L(\gamma,\underline q^2,\hslash) e^{i\vec {\underline q}\cdot \vec b}\,.
\end{equation}
from momentum-transfer space to impact-parameter space.
The classical eikonal phase $\delta(\gamma,b)$ is associated with an
exponentiation  of the $S$-matrix in $b$-space,
The effect is a representation of the amplitude in representations of terms that exponentiate with the eikonal phase $\delta(\gamma,b)$ and terms that do not and have to be incorporated in $\Delta(\gamma,b,\hslash)$
\begin{equation}\label{e:Texp}
  1+i\mathcal T=\left(1+i2\Delta(\gamma,b,\hslash)\right) e^{2i\delta(\gamma,b)\over\hslash}\,.
\end{equation}
The exponentiation of the amplitude in impact parameter space is a consequence of the unitarity relations of the $S$-matrix \cite{Cristofoli:2020uzm} and it has an analog to the method of Born subtractions.\\[10pt]
Furnished with amplitudes, we can regard the phase $\delta(\gamma,b)$ as independent of quantum corrections. It is conventional to expand it in powers of $G_N$ through perturbative orders where scattering amplitudes are expanded as $1+i\mathcal T = 1+i\sum_{L\geq0}  \mathcal
  M_L(\gamma,b,\hslash)$, thus we have
\begin{equation}
  \delta(\gamma,b)= \sum_{L\geq0} \delta_L(\gamma,b)\,,
\end{equation}
Using the eikonal phases carefully extracted from amplitudes we can assess observables such as the scattering angle by solving saddle-point conditions.
Utilizing this we can instantly infer at tree-level and one-loop level
\begin{equation}\label{e:delta1}
\delta_0(\gamma,b)=G_N m_1
  m_2\frac{ 2\gamma^2-1 }{ \sqrt{\gamma^2-1}}{(\pi b^2
e^{\gamma_E})^{4-D\over2}\over D-4}+\mathcal O((D-4)^0)\,,
\end{equation}
\begin{equation}
\delta_1(\gamma,b)=G_N^2 (m_1+m_2) m_1 m_2\frac{3 \pi (5\gamma^2-1)}{8 b \sqrt{\gamma^2-1}} (\pi b^2 e^{\gamma_E})^{4-D}+\mathcal O(4-D)\,,
\end{equation}
and if we take care of iterations at two-loop order
\begin{multline}\label{e:delta2}
\delta_2(\gamma,b)=\frac{G_N^3 m_1
  m_2 (\pi b^2e^{\gamma_E })^{3(4-D)\over2}}{2b^2
  \sqrt{\gamma^2-1}}\Bigg(\frac{2(12\gamma^4-10\gamma^2+1)\mathcal E_{\rm C.M.}^2}{\gamma^2-1}
\cr
-{4m_1m_2\gamma\over3}(25+14\gamma^2)+\frac{4 m_1 m_2(3+12
  \gamma^2-4\gamma^4)\arccosh(\gamma)}{\sqrt{\gamma^2-1}} \Bigg)\cr
+\frac{2 m_1 m_2(2\gamma^2-1)^2}{\sqrt{\gamma^2-1}}
{1\over
  (4(\gamma^2-1))^{4-D\over2}}\bigg(-\frac{11}{3}+\frac{d}{d\gamma}
\Big(\frac{(2\gamma^2-1)\arccosh(\gamma)}{\sqrt{\gamma^2-1}} \Big)
\bigg)\Bigg) +\mathcal O(4-D)\,.
\end{multline}
These results agree with ref. \cite{DiVecchia:2020ymx}.\\[5pt]
\section{New computational technology for gravitational scattering amplitudes}\label{sec:newmethods}
As we have observed we can compute scattering amplitudes from unitarity cuts using on-shell compact expressions for tree amplitudes. To streamline such computations concentrated on extracting the classical part from the full quantum amplitude it is advantageous to adjust the computational method and adapt computations according to a large mass
expansion. This has presently been accomplished by pursuing two routes. The first one is
a transformation of the heavy-mass effective theory
approach~\cite{Brandhuber:2021kpo,Brandhuber:2021eyq}, and the second one is to
adjust the unitarity method for determining the mass contributions from
massive cuts. This later technique dubbed the velocity cuts method
in~\cite{Bjerrum-Bohr:2021vuf,Bjerrum-Bohr:2021din,Bjerrum-Bohr:2021wwt}
is the one that we will demonstrate here. Such methods derive a refined computational path that concentrates on the integrand components in the amplitude that deliver the classical radial action. \\[5pt]
The starting point for this refined examination is the extraction of the radial action from an exponentiated $S$-matrix. As we have witnessed in the previous section the exponential phases $\delta$ in impact parameter space are constrained by the unitarity of
the $S$-matrix and they play a critical role in deriving the
classical Einstein's gravity. This motivates the following framework for computations~\cite{Damgaard:2021ipf} where an exponential representation of the $S$-matrix at the operator level in momentum transfer space was developed as a way to generate directly the radial action. We will in the subsequent review this formalism as it plays an essential role in the simplified computational framework that we will evolve later in the section. We begin with
\begin{equation}
  \widehat S= \mathbb I+{i\over\hslash}\widehat T= \exp\left(i \widehat N\over\hslash\right) ,
\end{equation}
where the scattering operator   $\hat T$ and the $\hat N$ operator (which can be associated with the radial action) have the expansion
\begin{equation}
  \hat T= \sum_{n\geq0} G_N^ n \hat T_n+ \sum_{n\geq0} G_N^{n+\frac12}
  \hat T^{\rm rad}_n; \qquad
  \hat N= \sum_{n\geq0} G_N^ n \hat N_n+ \sum_{n\geq0} G_N^{n+\frac12}
  \hat N^{\rm rad}_n;
  \end{equation}
together with the completeness relation
\begin{equation}\label{e:complet}
\mathbb I = \sum_{n=0}^{\infty} \frac{1}{n!} \int
\prod_{i=1}^2\frac{d^{D-1}k_i}{(2\pi\hslash)^{D-1}} \frac{1}{2E_{k_{1}}}
\prod_{j=1}^n 
\frac{d^{D-1}\ell_j}{(2\pi\hslash)^{D-1}} \frac{1}{2E_{\ell_{j}}}  
\left| k_1, k_2; \ell_1, \ldots \ell_n \right\rangle\left\langle k_1, k_2; \ell_1, \ldots \ell_n \right|,
\end{equation}
which encloses all the exchanges of gravitons for $n\geq1$ entering the
radiation-reaction contributions $\hat T^{\rm rad}$ and $\hat N^{\rm rad}$.
With this exponential representation of the $S$-matrix, we can
 systematically correlate matrix elements of the operator in the
 exponential $\hat N$ to ordinary Born amplitudes minus components
 supplied by unitarity cuts~\cite{Damgaard:2021ipf}. This is 
 witnessed by the perturbation expansion
 \begin{align}\label{e:NtoT}
 \hat{N}_0  &=  \hat{T}_0, \qquad
 \hat{N}_0^{\rm rad} = \hat{T}_0^{\rm rad}, \cr
\hat{N}_1  &=  \hat{T}_1 - \frac{i}{2\hslash}\hat{T}_0^2, \qquad
\hat{N}_1^{\rm rad} = \hat{T}_1^{\rm rad} -\frac{i}{2\hslash}(\hat{T}_0\hat{T}_0^{\rm rad} + \hat{T}_0^{\rm rad}\hat{T}_0) ,\cr
\hat{N}_2  &=  \hat{T}_2  - \frac{i}{2\hslash}(\hat{T}_0^{\rm rad})^2 - \frac{i}{2\hslash}(\hat{T}_0\hat{T}_1 + \hat{T}_1\hat{T}_0) - \frac{1}{3 \hslash^2}\hat{T}_0^3,
\end{align}
and similarly for higher orders.

\begin{figure}[h!!]
\centering \includegraphics[width=4.5cm]{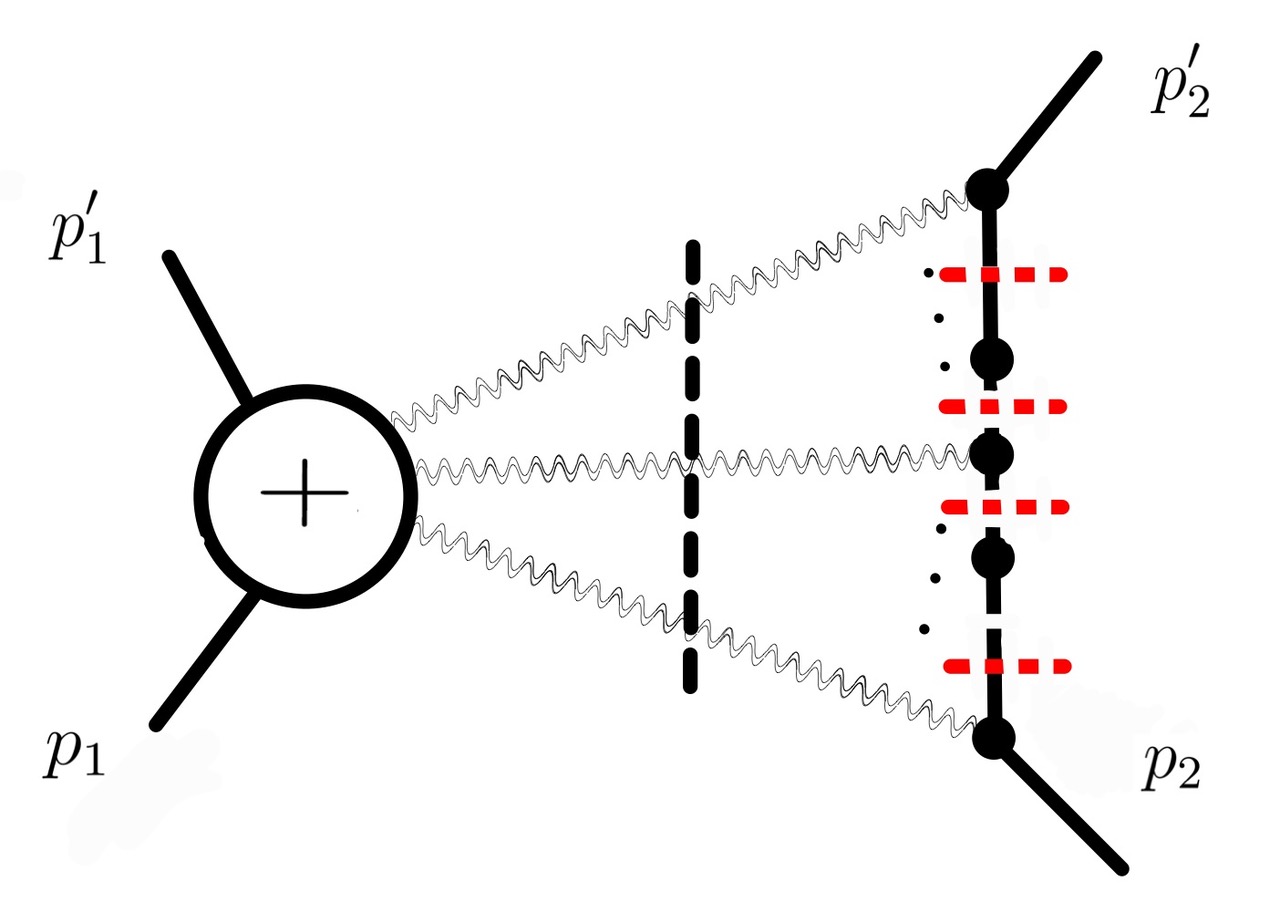} \!\!\!\! \!\!\!\! \!\!\!\!\includegraphics[width=3.5cm]{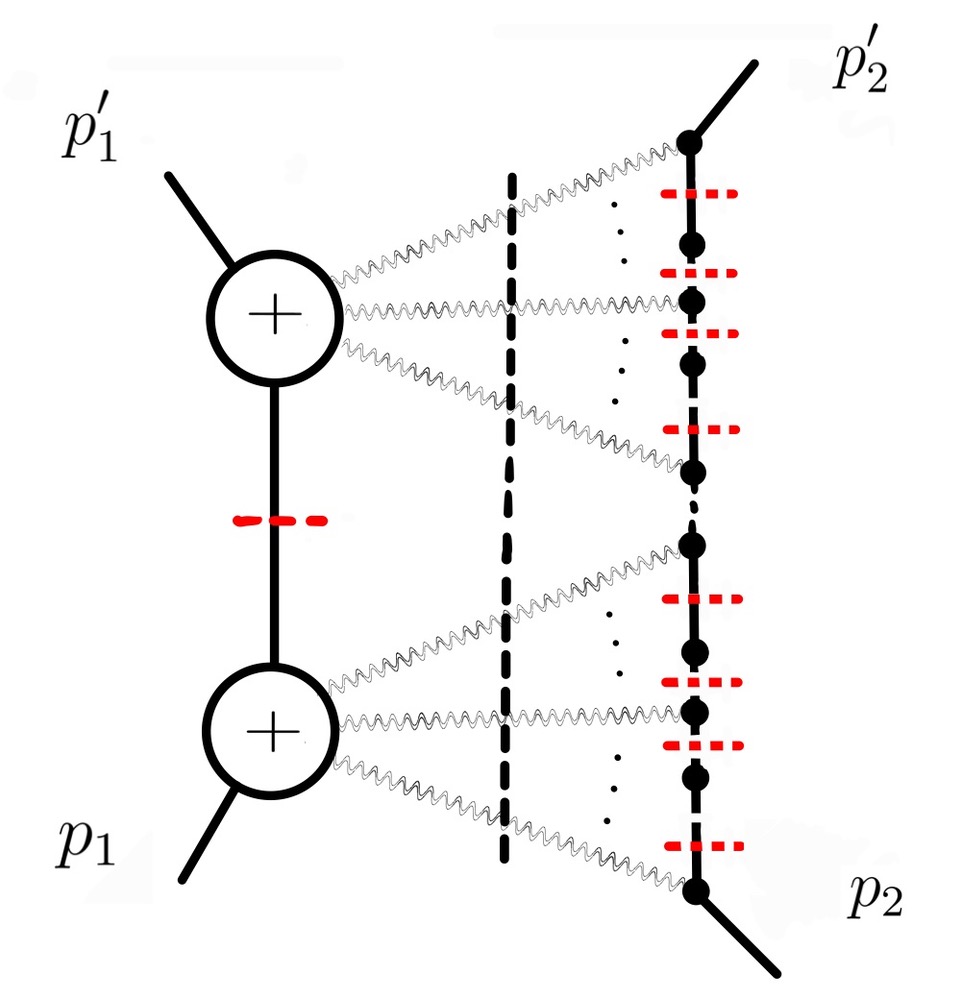}
\caption{The probe (left) and next-to-probe contribution (right).} \label{fig:probenextprobe}
\end{figure}

From the higher order terms in the $\hslash$ expansion~\eqref{e:NtoT},
one can as well study the  quantum
gravity  corrections~\cite{Bjerrum-Bohr:2002ks,Donoghue:2001qc,
 Bjerrum-Bohr:2014zsa, Brandhuber:2019qpg,AccettulliHuber:2019jqo,AccettulliHuber:2020dal} to the leading classical contributions, and
appeal to unitarity and analyticity arguments~\cite{Bellazzini:2015cra}
for constraining some higher-derivative terms in the effective action~\eqref{e:Leff}.
The next step for the exploration is multi-soft expansions of the tree-level amplitudes that enter unitarity cuts in the integrand construction. First, we determine a basis of integral functions that enter a given unitarity cut, next we consider the constraints levied by unitarity on the integrand. We can for instance at $L$ loop order consider a generalized $(L+1)$ graviton cut.\footnote{\footnotesize{The multi-graviton cut is not enough to for reconstructing the full classical $L$-loop amplitude. At two-loops, we for instance require to add the bowtie diagram where the amplitude is factorized into a product of two scalar-graviton amplitudes times graviton amplitude. Another type of graph we must add is self-energy graphs}}
\begin{multline}\label{e:MLcut}
 i\mathcal M_{L}^{\textrm{cut}} (\sigma,q^2)=\hslash^{3L+1} \int (2\pi)^D\delta(q+\ell_2+\cdots+\ell_{L+2})\prod_{i=2}^{L+2} {i
 \over\ell_i^2}\prod_{i=2}^{L+2} { d^D\ell_i\over (2\hslash\pi)^D} 
 \cr{1\over (L+1)!}\sum_{h_i=\pm2}
 M^{\rm tree}_{\rm Left}(p_1,\ell_2^{h_2},\ldots,\ell_{L+2}^{h_{L+2}},-p_1')
 M^{\rm tree}_{\rm Right}(p_2,-\ell_2^{h_2},\ldots,-\ell_{L+2}^{h_{L+2}},-p_2')^\dagger,
\end{multline}
Here $M^{\rm tree}_{\rm Left}(p_1,\ell_2,\ldots,\ell_{L+2},-p_1')$ and
$M^{\rm tree}_{\rm Right}(p_2,-\ell_2,\ldots,-\ell_{L+2},-p_2')$ are tree-level
amplitudes corresponding to multi-graviton emission from a massive scalar line
Now in all outgoing line conventions, we can write the following equation for momentum conservation
\begin{equation}
 q=p_1-p_1'=-\sum_{i=2}^{L+2}\ell_i. 
\end{equation}
Now an important observation is that if we consider a tree amplitude as
\begin{equation}\label{e:MLhat}
 M^{\rm tree}_{\rm Left}(p_1,\ell_2,\dots,\hat\ell_i,\ldots,\ell_{L+1},{\ell}_{L+2},-p_1'),
\end{equation}
where we use momentum conservation
\begin{equation}
 \ell_i=-q-\sum_{2\leq j\leq L+2\atop j\neq i}\ell_j.
\end{equation}
One observes that in~\eqref{e:MLhat} there exist different types of massive
propagators. We have propagators with a `hatted' momentum for which we can write
\begin{equation}
\frac{1}{(p_1-\ell_{i_2}-\cdots-\ell_{i_j}-q)^2-m^2+i \varepsilon}=\frac{-1}{2p_1\cdot(\ell_{i_2}+\cdots+\ell_{i_j}) -(\ell_{i_2}+\cdots+\ell_{i_j}+q)^2-i \varepsilon},
\end{equation}
and propagators without a `hatted' momentum
\begin{equation}
\frac{1}{(p_1+\ell_{i_2}+\cdots+\ell_{i_j})^2-m^2+i \varepsilon}=\frac{1}{2p_1\cdot(\ell_{i_2}+\cdots+\ell_{i_j}) +(\ell_{i_2}+\cdots+\ell_{i_j})^2+i \varepsilon}\,.
\end{equation}
(we have taken $1<j\leq L+2$).\\[5pt]
Now the important observation is the following. If we employ the equation
\begin{equation}\label{e:PP}
 \lim_{\varepsilon\to0^+}\left( {1\over \eta-i\varepsilon}-{1\over \eta+i\varepsilon}\right)=
 \lim_{\varepsilon\to0^+} {2i\varepsilon \over \eta^2+\varepsilon^2}=2i\pi
\delta(\eta),
\end{equation}
and recast propagators with a `hat'  
\begin{multline}\label{e:propflip}
\frac{1}{(p_1-\ell_{i_2}-\cdots-\ell_{i_j}-q)^2-m^2+i \varepsilon}=
-2i\pi \delta\big((p_1-\ell_{i_2}-\cdots-\ell_{i_j}-q)^2-m^2\big)\cr
+\frac{1}{(p_1-\ell_{i_2}-\cdots-\ell_{i_j}-q)^2-m^2-i \varepsilon}\,.
\end{multline}
we are directed to a new organizing principle for integrands. Iterating applications of this identity one can derive a form tree amplitudes with two scalar legs and $L+1$ graviton legs organized with different powers
of unitarity cuts delta functions. Defining 
$M^{\textrm{tree}(+)}_{L+1}(p_1,\ell_2, \dots,{\ell}_{L+2},-p_1')$ as the tree-level amplitudes with propagators $2p_1\cdot (\sum_r \ell_r)- (q+\sum_r \ell_r)^2-i
\varepsilon$ rewritten as $2p_1\cdot (\sum_r \ell_r)- (q+\sum_r
\ell_r)^2+i \varepsilon$,
and similarly $M^{\textrm{tree}(-)}_{L+1}(p_1,\ell_2,\dots,{\ell}_{L+2},-p_1')$ with propagators $2p_1\cdot (\sum_r \ell_r)+ (\sum_r \ell_r)^2+i \varepsilon$ changed to $2p_1\cdot (\sum_r \ell_r)+(\sum_r \ell_r)^2-i \varepsilon$,
we arrive at
\begin{multline}\label{e:MLdelta}
M^{\rm tree}_{L+1} \sim (M^{\textrm{tree}(+)}_1)^{L+1} \prod_i^L
\delta_i(\ldots) +(M^{\textrm{tree}(+)}_1)^{L-1}
(M^{\textrm{tree}(+)}_2) \prod_i^{L-1} \delta_i(\ldots)+\cdots\cr
+M^{\textrm{tree}(+)}_1 M^{\textrm{tree}(+)}_L \delta(\ldots)+M^{\textrm{tree}(+)}_{L+1}\,,
\end{multline}
as a new representation of the tree-level amplitude.
The important observation is that it is now relatively easy to read the soft behavior of such trees in unitarity cuts, since considering the soft scaling behavior we have 
\begin{equation}
 \delta{\left((p_1+ \sum \ell_i)^2-m_1^2\right)} = \delta \left(2|q| p_1\cdot \sum \tilde\ell_i+\mathcal O(|q|^2)\right)={1\over|q|} \delta\left(2 p_1\cdot \sum \tilde\ell_i\right)+\mathcal O(|q|^0),
\end{equation}
and thus the amplitude $\mathcal M_{L+1}$ has multi-soft
scaling 
\begin{equation}\label{e:Treemultisoft}
\lim_{|\vec q|\to0} M^{\rm tree}_{L+1}(p,|\vec q|\tilde\ell_2,\dots,|\vec
q|\hat{\tilde\ell}_{L+2},-p')
=\frac{(M^{\rm tree}_{1})^{L+1}\delta(\cdots)^{L}}{|\vec{q}|^{L}}+ \mathcal
O\left(1\over |q|^{L-1}\right)\,.
\end{equation}
Using this decomposition in integrand derived from generalized unitarity and combining this knowledge with the exponential representation of the $S$-matrix
provided in ref.~\cite{Damgaard:2021ipf} one arrives simply with an integrand of the classical radial action. This was in ref.\cite{Bjerrum-Bohr:2021wwt} developed as a new practical technique for calculating post-Minkowskian dynamics.\\[5pt]
For instance, focusing on the leading probe contribution we see that it arises the following type of contribution in the cut
\begin{multline}
(M_{\rm Left} {M_{\rm Right}}^\dagger)\Big|_{\rm
  probe}=
M^{\textrm{tree}(-)\dagger}_{L+1}(p_2,-\hat{\ell}_2,-\ell_3,\ldots,-\ell_{L+1},-p_2')\cr
\times M^{\textrm{tree}(+)}_1(p_1,\hat{\ell}_2,\!-p_{1}\!-\!\hat{\ell}_2)\prod_{j=3}^{L}\delta((p_1+\hat{\ell_2}+
\cdots +\ell_{j-1})^2-m_1^2)\cr
M^{\textrm{tree}(+)}_1(p_{1}+\hat{\ell}_2+ \cdots
+\ell_{j-1},\ell_j,\!-\!p_{1}\!-\!\hat{\ell}_2- \cdots +\ell_j),
\end{multline}
where the cut is evaluated using $\ell_i^2=0$ for $1\leq i\leq L+1$.
One can represent this contribution as in the left figure~\ref{fig:probenextprobe}.
Contributions also arise from next-to-probe topologies which can be
represented in the right figure~\ref{fig:probenextprobe}.\\[-30pt] %
%
\section{Fourth-order post-Minkowskian dynamics}\label{sec:4PM}
At the fourth post-Minkowskian order, new puzzles arise regarding the
derivation of gravitational radiation and this presents a strong test for
validating the understanding of the logic developed
for deriving the post-Minkowskian expansion in classical gravity.

At this order in perturbation, the contributions can be split into 
conservative and radiation pieces, the conservative has been derived  in~\cite{Bern:2021dqo,Bern:2021yeh, Bern:2022jvn, Bini:2021gat,Dlapa:2021npj,Dlapa:2022lmu}. 
In this section, we explain that the application of  the velocity cut
formalism, presented in the previous section, leads to re-derivation
of  the conservative piece from a small subset of maximal cut
topologies. A complete understanding of the radiative corrections is still an open
question. We will present a partial derivation of these radiative corrections
and comment on the current puzzles associated with such contributions.

The real part of the classical two-body gravitational interaction
from the three-loop scattering  has the mass expansion
\begin{multline}\label{e:N4PMlim}
    \mathcal M_{\rm 4PM}(\gamma,\underline q^2)|_{\rm classical}=\lim_{\epsilon\to0}\Bigg({G_N^4
      m_1^2m_2^2\over|\underline q|^{-1+3\epsilon}}(m_1+m_2) \Big(
  (m_1+m_2)^2 c^{\rm probe}(\gamma,\epsilon) \cr+m_1m_2  c^{\rm next-to-probe}(\gamma,\epsilon)\Big)\Bigg).
\end{multline}
The amplitude is evaluated in $D=4-2\epsilon$, for control of the
infra-red divergences from gravitational radiation. The expression has
a finite four-dimensional limit as expected. This result has the generic
polynomial mass dependence from~\eqref{e:MLclassical} with $L=4$ as a consequence of the
delta-functions counting for obtaining the classical piece.

\begin{figure}[h]   \centering
\!\!\hskip-0.5cm$ \hskip1cm
\lim_{\hslash\to0}
 \begin{gathered}
    \includegraphics[width=1.3cm]{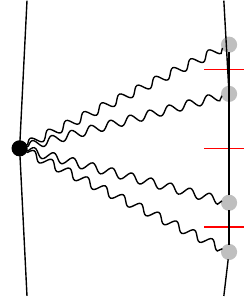}
     \end{gathered}=  \begin{gathered}
       \includegraphics[width=1.3cm]{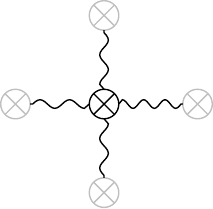}\end{gathered},
     \ \ 
\lim_{\hslash\to0}   \begin{gathered}
   \includegraphics[width=1.3cm]{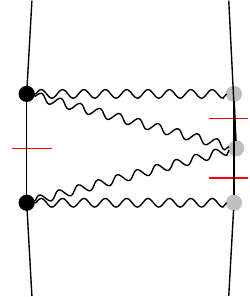}
        \end{gathered} =
        \includegraphics[width=1.3cm]{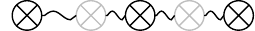},\ \ 
 \lim_{\hslash\to0}
       \begin{gathered}
         \includegraphics[width=1.3cm]{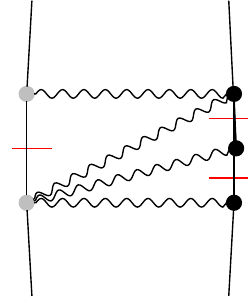}
      \end{gathered} =
            \begin{gathered}
        \includegraphics[width=1cm]{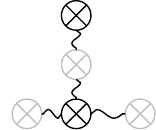}
            \end{gathered}
$
    \hskip-1.7cm$ \hskip1cm
     \lim_{\hslash\to0}  \begin{gathered}
   \includegraphics[width=1.4cm]{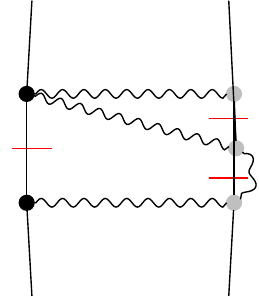}
     \end{gathered}=  \begin{gathered} \includegraphics[width=1.5cm]{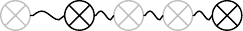} \end{gathered}, \qquad
     \lim_{\hslash\to0}\begin{gathered}
   \includegraphics[width=1.4cm]{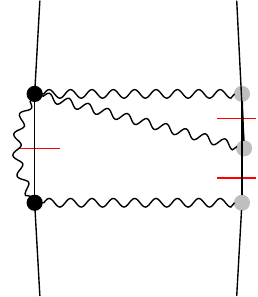}
        \end{gathered}  =
        \includegraphics[width=1cm]{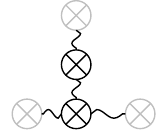}$,\\
$  \lim_{\hslash\to0}\begin{gathered}
  \includegraphics[width=1.4cm]{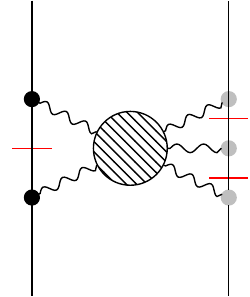}
        \end{gathered} =
                \begin{gathered}
              \includegraphics[width=1.4cm]{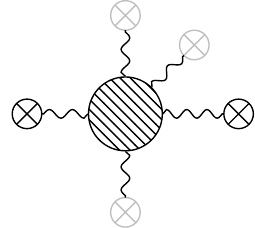}
                \end{gathered},\qquad 
     \lim_{\hslash\to0}\begin{gathered}\includegraphics[width=1.5cm]{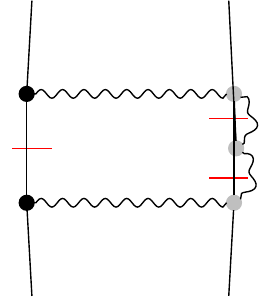}\end{gathered}
  =  \begin{gathered}\includegraphics[width=1.5cm]{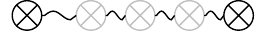}\end{gathered}$
\caption{The classical limit of the velocity cuts.}    \label{fig:4cuttoPlanar}

      \end{figure}

Taking the classical limit $\hslash\to0$ imposes delta-function
insertions massive lines as explained in the previous section.
They are represented by a red line on 
in the figure~\ref {fig:4cuttoPlanar}. 
These delta-function insertions cut
open the massive lines and lead to the 
important simplification of reducing the integrand
at the fourth order post-Minkowskian order
to a sum over permutations of planar
tree graphs listed in fig.~\ref{fig:tree}. 

\begin{figure}[h!!]
\centering  \includegraphics[width=11cm]{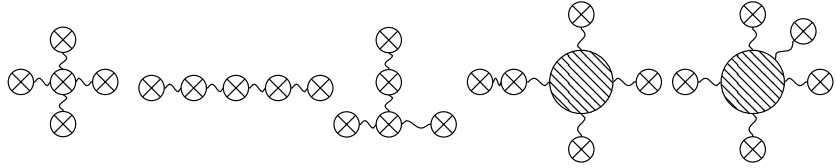}
\caption{There are only four different tree topologies needed in the reconstruction of the full 4 PM integrand. We
denote sources by crossed dots in the graphs, the dashed blob is a multi-graviton amplitude.}
\label{fig:tree}
\end{figure}

In assembling the integrand is the fact that
non-planar integrand topologies can be appearing~\cite{Bern:2021dqo,Bern:2021yeh, Bern:2022jvn, Dlapa:2021npj, Dlapa:2022lmu}. However, because of
the velocity cuts, we have found that is always possible to assemble
these contributions by use of a partial fractioning identity that maps
the non-planar integral denominator into a combination of planar
denominators using that when the scalar line is cut open permutations
of the graviton lines attached to the scalar lines are mapped to the
same type tree-level graph topologies above modulo a permutation of
the assignment of legs. The non-planar contributions from the
multi-graviton tree-level insertions are thus rephrased in the terms
of planar removed and thus as a consequence, the reduction of master
integrals need only involve planar graphs. Another technical point is
the fact that overlapping cut topologies need to be eliminated between
cut topologies. We eliminate such overlaps directly in our
reconstruction of the integrand by the introduction of symmetry
factors in certain integrand sectors. For instance the
  contribution with a four-graviton amplitude in the figure below is already accounted for in the listed topologies
$$
    \includegraphics[]{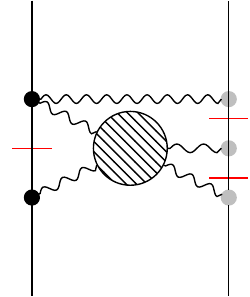}   
$$
Performing the tensor reduction with {\tt
 LiteRed}~\cite{Lee:2013mka} the fourth post-Minkowskian radial
action is then expanded on forty master integrals
\begin{equation}
  \mathcal M_{\rm 4PM}(\gamma,\underline q^2)=\lim_{\epsilon\to0} \sum_{i=1}^{40}
  c\left(\{n_j\};\gamma,\underline q^2\right)  \mathcal I\left(\{n_j\};\gamma,\epsilon\right)\,,
\end{equation}
with the family of master  integrals evaluated in $D=4-2\epsilon$
dimensions
\begin{equation}\label{e:MasterDef}
  \mathcal I\left(\{n_j\};\gamma,\epsilon\right)=
  \int \frac{1}{\prod_{i=1}^{12}D_i^{n_i}}\prod_{r=1}^3 {2\pi i
    \delta(2u_1 \cdot\sum_{j=1}^r l_j) d^{4-2\epsilon}l_i\over (2\pi)^{4-2\epsilon}}\,,
\end{equation}
where we have defined  the propagators
 \begin{align}\label{e:PropDefNew}
&D_1=(l_1+l_2+q)^2, \  D_2=(l_1+l_2)^2, \  D_3=2 u_1 \cdot (l_1+l_2),\ 
       D_4=l_2^2,\   D_5=l_3^2, \cr
  &  D_6=l_1^2,   D_7=(l_1+l_2+l_3+q)^2 , D_8=(l_1+q)^2,  D_9=(l_1+l_2+l_3)^2,\cr
&   D_{10}=-2u_2 \cdot l_1 ,
  D_{11}=-2u_2 \cdot (l_1+l_2+l_3), \   D_{12}=(l_2+l_3)^2\,,
\end{align}
with  
$u_1^2=u_2^2=1$, $\gamma=u_1\cdot
u_2$, $q^2=-1$, $u_1\cdot q=u_2\cdot q=0$.  It is useful to introduce the vector
$k_1^2=-1$, such that $u_1\cdot k_1=0$ and $u_2=\gamma u_1+\sqrt{\gamma^2-1}
k_1$.
Since the integrals will only have infrared
  divergences we take $\epsilon<0$ (we refer to~\cite{Bjerrum-Bohr:2021vuf}
  for a discussion of the infrared divergences in classical gravity computations).
The set of master integrals in~\eqref{e:MasterDef} have three
delta-functions from the velocity cuts on the massive propagator as required for obtaining classical piece of order
$1/\hslash$ from the quantum
amplitude~\cite{Bjerrum-Bohr:2021wwt}. This is the 
same class of Post-Minkowskian master integrals as the one
obtained  by the word-line EFT
approach of~\cite{Kalin:2020fhe,Dlapa:2021npj,Jinno:2022sbr}.

Changing variables to  the rapidity
\begin{equation}\gamma=\frac12\left(x+{1\over x}\right),
\end{equation}
the forty master integrals satisfy the linear differential
system
\begin{equation}
  {d\over dx}   \vec {\mathcal I}(x,\epsilon)= A(x,\epsilon)  \vec {\mathcal I}(x,\epsilon)\,,
\end{equation}
with (regular) singularities at $x=\{-1, +1\}$ corresponding to the
static limit $\gamma^2=1$ and at $x=\{0, \infty\}$ corresponding to the
high-energy regime $\gamma\to\infty$. There are as well
apparent\footnote{Apparent singularities are singularities of
  $A(x,\epsilon)$ where the general solutions $\vec {\mathcal I}(x,\epsilon)$
  are holomorphic. The removal of the apparent singularities can be
  done using {\tt Libra}~\cite{Lee:2020zfb} or using the {\tt
    ore\_algebra} package implementation in~\cite{ore,ore2}.}
   singularities at the roots of $x^4 \left(416 \epsilon ^3-354
     \epsilon ^2+91 \epsilon -6\right)\ +x^2 \left(320 \epsilon
   ^3-292 \epsilon ^2+82 \epsilon -6\right)+416 \epsilon ^3-354 \epsilon ^2+91 \epsilon
 -6=0$.   At this order in perturbation arises an elliptic sector
 with solution are product of elliptic functions~\cite{Bern:2021yeh,Dlapa:2021npj}.

Using the basis of master integrals as in the probe computation
in~\cite{Bjerrum-Bohr:2021wwt}, we can expand all the forty master
integrals of the basis in the potential region, giving boundary data.
This is enough to
obtain the full solution of the differential system in terms of only 
planar integrals.
By matching the post-Newtonian expansion, we find that  the system is decomposed into six sectors depending on the regime of the
 loop momenta $l_2$ and $l_3$ of the master integrals
 in~\eqref{e:MasterDef}. There are three regions $I_{\rm PP}^i(\epsilon)$ where the loop momenta
$l_2$ and $l_3$ correspond to graviton 
on potential modes,  two regions $I_{\rm PR}^i(\epsilon)$ where on loop momentum
$l_2$ is potential modes  and the loop $l_3$ momentum is on radiation
mode, and a region  $I_{\rm RR}(\epsilon)$ where the loop momenta
$l_2$ and $l_3$ correspond to graviton 
on a radiation mode. 

 The piece of the amplitude proportional to the  mass dependence $m_1^5m_2^2$
and $m_1^2 m_2^5$ is the 
probe contribution given by the first four-graviton cut  in fig.~\ref{fig:4cuttoPlanar}.
The expression was derived
in eqs~(5.19),~(5.18) of~\cite{Bjerrum-Bohr:2021wwt} with the result
\begin{align} 
   \mathcal M_{\rm 4PM}^{\rm probe}(\gamma,\underline q^2)   &=\lim_{\epsilon\to0} {(8\pi G_N)^4 \over |\underline q|^{-1+3\epsilon}} m_1^2m_2^2(m_1^3+m_2^3)
   {(1-2\epsilon)^3\over (2-2\epsilon)^4}
   {c_3(\gamma,\epsilon)\over (\gamma^2-1)^3} I_{\rm
                                                               PP}^1(1,\epsilon),\cr
                                                               &=
   G_N^4(m_1^3+m_2^3)m_1^2m_2^2|\underline q|\pi^3  \frac{35 i \left(33 \gamma ^4-18 \gamma ^2+1\right)}{8 \left(\gamma
   ^2-1\right)}\,,\label{e:Probe}
 \end{align}
 with the coefficient $c_3(\gamma,\epsilon)$ evaluated in eq.~(5.17)
of~\cite{Bjerrum-Bohr:2021wwt} and  the master  integral $I_{\rm
  PP}^1(\epsilon)$ 
is the massless sunset integral evaluated in eq.~(2.31)
of~\cite{Mougiakakos:2020laz} . This is the probe contribution denoted $ \mathcal
M_4^p(\gamma) $ in~\cite{Bern:2021yeh}.

The next-to-probe contribution to the radial action $\mathcal N^{\rm next-to-probe}_{\rm 4PM}(\gamma)$ is exacted from the mass
 contributions  $m_1^4m_2^3$  or $m_1^3m_2^4$ of the three-loop
 amplitude, arising from the graphs with one
velocity cut on the left  and two velocity cuts on the right in the graphs~\ref{fig:4cuttoPlanar}
 \begin{equation}
\mathcal M_{\rm 4PM}^{\rm next-to-probe}(\gamma,\underline q^2)=-3  \mathcal M_{\rm 4PM}^{\rm probe}(\gamma,\underline q^2) +\lim_{\epsilon\to0} {G_N^4
 m_1^3m_2^3  (m_1+m_2)\over |\underline
 q|^{-1+3\epsilon} } \widehat{\mathcal M}_3(\gamma,\epsilon)\,,
\end{equation}
with the expansion on the static master integrals
\begin{multline}\label{e:nextotprobe}
  \widehat{\mathcal M}_3(\gamma,\epsilon)={\pi^2\over2} c_{\rm PP}^2(\gamma)
    I_{\rm PP}^2(\epsilon)
   +\sum_{i=1,3} c_{\rm PP}^i(\gamma, \epsilon)
    I_{\rm PP}^i (\epsilon)\cr+\left(4(\gamma^2-1)\right)^{-2\epsilon} c_{\rm RR}(\gamma, \epsilon)
    I_{\rm RR}(\epsilon) + \left(4(\gamma^2-1)\right)^{-\epsilon} \sum_{i=1}^2 c^i_{\rm PR}(\gamma)
    I_{\rm PR}^i (\epsilon).
  \end{multline}
  The four dimensional limit, $\epsilon\to0$, is infrared finite and
splits into two contributions
\begin{equation}
  \lim_{\epsilon\to 0}   \widehat{\mathcal M}_3(\gamma,\underline q^2,\epsilon)=\widehat{\mathcal M}_3^{\rm PP-RR}(\gamma,\underline
  q^2)+ \widehat{\mathcal M}_3^{\rm PR}(\gamma,\underline
  q^2).
\end{equation}
A first contributions from the potential-potential and the
radiation-reaction regions
\begin{align}\label{e:nextotprobePRRR}
  \widehat{\mathcal M}_{3} ^{\rm
    PP-RR} (\gamma,\underline q^2)&= \lim_{\epsilon\to0}\Big(\sum_{i=1,3} c_{\rm PP}^i(\gamma, \epsilon)
    I_{\rm PP}^i (\epsilon)+{c_{\rm RR}(\gamma, \epsilon)\over\left(4(\gamma^2-1)\right)^{2\epsilon}} 
                                    I_{\rm RR}(\epsilon)\Big)\\
  &+{\pi^2\over2} c_{\rm PP}^2(\gamma)
    I_{\rm PP}^2(0),\cr
                  \nonumber                 & =
  -16i\left( 4 \mathcal M_4^t(\gamma) \log\left(\sqrt{\gamma^2-1}\over2\right)+ \mathcal
  M_4^{\pi^2}(\gamma) + \mathcal M_4^{\rm rem} (\gamma) +3 \mathcal M_4^p(\gamma) \right),
\end{align}
with the tail contribution
\begin{multline}
-64i \mathcal M_4^t  (\gamma)= -\frac{16 i \gamma  \left(2 \gamma
      ^2-3\right) \left(35 \gamma ^4-30 \gamma
      ^2+11\right)}{\left(\gamma ^2-1\right)^{3/2}} \arccosh(\gamma)\cr
 +32 i \left(35 \gamma ^4+60 \gamma ^3-150 \gamma ^2+76 \gamma -5\right) \log
  \left(1+\gamma\over2\right).
\end{multline}
The expression~\eqref{e:nextotprobePRRR} from the potential-potential  and the
radiation-reaction regions  reproduces the conservative part derived in~\cite{Bern:2021yeh}.
The radiation-radiation sector cancels the infra-red divergence from
the potential-potential region leading to an infrared
finite result for the classical result~\cite{Bern:2021yeh}.

The contribution from the potential-radiation region reads
\begin{align}\label{e:nextotprobePR}
  \widehat{\mathcal M}^{\rm
    PR}_{3}(\gamma)&=\lim_{\epsilon\to0}\left(4(\gamma^2-1)\right)^{-\epsilon} \sum_{i=1}^2 c^i_{\rm PR}(\gamma)
    I_{\rm PR}^i (\epsilon)\cr
                     &=
  { 3\pi\over2}   (5 \gamma
                                        ^2-1)  {2\gamma^2-1\over\sqrt{\gamma^2-1}}\left( {\gamma ^2\over\gamma^2-1}+\frac{2
   \left(2 \gamma ^2-3\right) \gamma
       \arccosh\left(\gamma\right)}{(\gamma
                     ^2-1)^{3\over2}}-\frac{16}{3}\right) \cr
            &         - \pi \, \mathcal M_4^t {d\over d\gamma}\left( {2\gamma^2-1\over\sqrt{\gamma^2-1}}\right).
\end{align}
The implication for the scattering angle is discussed in
section~\ref{sec:observables}.

We conclude this section by stressing that 
a complete derivation of the fourth post-Minkowskian contribution from a scattering amplitude
approach is unfortunately not yet fully understood.
One signal that the previous analysis is incomplete is that the
high-energy $\gamma\to\infty$ the scattering amplitude does not
reproduce the expected 
behaviour of $\gamma^2$ dominated by the probe contribution according~(7.1)
of~\cite{Damour:2017zjx}. Actually all
the results derived in the
literature~\cite{Bern:2021yeh,Dlapa:2021npj,Dlapa:2022lmu}, at the
time of writing this text,
fail to reproduce this  behaviour.

The extraction of the classical piece is very similar to the large-mass
expansion~\cite{smirnov1,Smirnov:1994tg} which requires a lot of care
in identifying the various asymptotic
regions~\cite{Gardi:2022khw}. The present analysis seems to have
identified all the 
regions contributing to the classical result some subtle contribution
could resolve the high-energy behaviour problem.
A complete understanding of the
problem of the gravitational reaction at this order in perturbation is
still an open question.

\section{Black-hole metrics}
\label{sec:schw}
The Nobel prize citation for Roger Penrose states that ``black hole formation is a robust prediction of the general theory
    of relativity''. Subrahmanyan Chandrasekhar explained that they are the most perfect macroscopic objects there
    are in the universe since the
only elements in their construction are our concepts of space and
time. Black-hole solutions are a perfect playground to validate
the formalism of deriving classical gravity from quantum scattering amplitudes. This
 also opens new avenues for studying black holes in generalized
theories of gravity.

By evaluating the vertex function of the emission of a graviton from a
particle of mass $m$,  spin $S$ and charge $Q$, in $d$ dimensions
\begin{equation}
\begin{gathered}
  \includegraphics[]{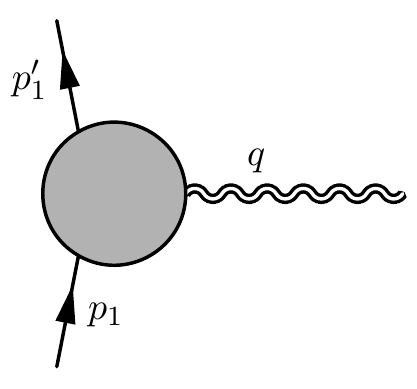}
\end{gathered}=-{i\sqrt{32\pi G_N}\over2}
\sum_{l\geq0}\langle  T^{(L)\, \mu\nu}(q^2)  \rangle\epsilon_{\mu\nu}
\end{equation}
At each loop order, we  extract the $L$-loop contribution to the
transition density of the stress-energy tensor
$\langle T_{\mu\nu}(q^2)\rangle=\sum_{l\geq0} \langle T^{(L)}_{\mu\nu}(q^2)\rangle$
\begin{equation}\label{e:MtoT}
  i\mathcal M^{ (L )}_3(p_1,q)  =-{i\sqrt{32\pi G_N}\over2}
\langle  T^{(L)\, \mu\nu}(q^2)  \rangle\epsilon_{\mu\nu},
\end{equation}
where $\epsilon^{\mu\nu}$ is the polarization of the graviton with
momentum $q=p_1-p_2$ is the momentum transfer. We also define $d\equiv D-1$.
 The de Donder gauge relation between the
metric perturbation $g_{\mu\nu}=\eta_{\mu\nu} +\sum_{n\geq1}
h^{(n)}_{\mu\nu}$ and the stress-energy tensor reads
\begin{equation}\label{e:TtohAmplitudedeDonder}
 h^{(L+1)}_{\mu\nu}(\vec x) = -16\pi G_N\int {d^d{\vec q}\over(2\pi)^d} e^{i\vec q\cdot \vec x} {1\over
 \vec q^2} \left( \langle T_{\mu\nu}^{(L)}\rangle^{\rm
   class.}(q^2)-\frac{1}{d-1}\eta_{\mu\nu}\langle T^{(L)}\rangle^{\rm class.}(q^2)\right)\,.
\end{equation}
In this relation enters the classical contribution at $l$ loop order $\langle T^{(L)}_{\mu\nu}\rangle^{\rm class.}(q^2) $ defined
by the classical limit of the quantum scattering
amplitude~\cite{Holstein:2004dn,Bjerrum-Bohr:2018xdl,Kosower:2018adc}.

With the action
\begin{equation}
\mathcal{S}=\int d^{d+1}x \sqrt{-g}\left({R\over 16\pi G_N}+
\frac{1}{2} g^{\mu\nu}\partial_{\mu}\phi \partial_{\nu}\phi-\frac{1}{2}m^2\phi^2\right)\,.
\end{equation}
one can extract the metric of   physical black holes
\begin{itemize}
\item Schwarzschild black hole: Scalar field $S=0$, mass
  $m$~\cite{Neill:2013wsa,Goldberger:2004jt,Bjerrum-Bohr:2002ks,Mougiakakos:2020laz}.  In~\cite{Mougiakakos:2020laz}, the Schwarzschild metric up to $G_N^4$
obtained in four ($d=3$), five ($d=4$) and six ($d=5$) dimensions.
  \item Reissner-Nordstr\"om black hole: Scalar field  $S=0$, charge
    $Q$, mass $m$~\cite{Donoghue:2001qc}
 \item Kerr-Newman  black hole: Fermionic field  $S=\frac12$, charge
    $Q$, mass $m$~\cite{Bjerrum-Bohr:2002ks,Donoghue:2001qc}
\end{itemize}
The derivation of the metric solutions in four dimensions from the scattering amplitude approach is a consistency
check on the way classical Einstein's gravity is embedded into the
standard massless spin-2 quantization of the gravitational
interactions. 
But, this can be applied to 
 any effective field theories of gravity and derive a black hole metric. This is particularly
interesting since black holes are expected to have interesting
behavior in the higher-dimensional theory of gravity~\cite{Emparan:2008eg,Emparan:2013moa}.

In the same way, the classical contribution was extracted from the
amplitude computation, one can in the effective field theory approach to gravity start processing quantum corrections in the amplitude. We will not delve into this here since it requires a refined partial wave expansion to interpret such terms correctly. We will however emphasize that unique $\hslash$ dependent contributions were computed in~\cite{Bjerrum-Bohr:2002ks,Donoghue:2001qc}, and the effective field approach opens up possibilities for investigating further low-energy consequences of quantum contributions to scattering amplitudes.

\section{Observables in General Relativity from amplitudes}\label{sec:observables}
Because in General Relativity the gauge transformation is as well a
coordinate frame change, the notion of observable differs from other
quantum field theories, say electromagnetism where any gauge invariant
expression is defined at a given time and position.

In the context of the scattering amplitude approach, it is natural to
connect to the scattering angle and the Shapiro time
delay/advance, which are some of the fundamental classical observables
in General Relativity. The scattering angle plays an important role in
connecting the various computations done in the post-Newtonian and
post-Minkowskian computations. The scattering angle is a central
quantity for addressing the question of the radiation and the
ultra-relativistic regime. The Shapiro time delay is used for
constraining various effective field theories of gravity~\cite{Hou:2017cjy,AccettulliHuber:2020oou,Dyadina:2021paa,Bellazzini:2021shn}.

By applying the same framework to the gravitational scattering of a massless
particle against a massive source, one can derive the second
post-Minkowskian correction to the  bending of light~\cite{Bjerrum-Bohr:2014zsa}
\begin{equation}\label{e:theta}
\theta  \simeq {4G_N M\over b}+{15\over 4} {G_N^2 M^2 \pi \over
  b^2}
+\ldots\,.
\end{equation}
We see that the eikonal  approximation leads to the expected
classical General Relativity contributions, in agreement  with
 the next-to-leading correction of~\cite{Akhoury:2013yua} and
 \cite{D'Appollonio:2010ae}. The contribution from John Donoghue in
 this volume~\cite{Donoghue:2022eay} discussed  the implication of
 this computation in the  framework of effective field theory of quantum gravity.

There are other observable like the impulse which  measures  the net
change in the momentum of one of the scattered particles in the
initial state which can be used to compute the classical interaction
potential and the eikonal phase~\cite{Kosower:2018adc,Kosower:2022yvp}. 

By scattering two massive spinless scalars one can 
the scattering angle at the first and second
post-Minkowskian order 

\begin{align}
  \chi_{\rm 1PM}&={2(2\gamma^2-1)\over\sqrt{\gamma^2-1}}\, {G_Nm_1m_2\over J},\cr
\chi_{\rm 2PM}&={3\pi \over
            4}{ m_1+m_2\over
            \mathcal E_{\rm C.M.}} (5\gamma^2-1)\left(
            G_Nm_1m_2\over J\right)^2\,.
\end{align}
where $\mathcal E_{\rm C.M.}^2=m_1^2+m_2^2+2m_1m_2\gamma$ and $J$ is the  angular momentum 
\begin{equation}\label{e:Jdef}
  J={m_1m_2\sqrt{\gamma^2-1}\over\mathcal E_{\rm C.M.}}b\cos\left(\chi\over2\right)\,.
\end{equation}
To this order, the scattering angle is the same as the one for the scattering of a test particle of mass 
$m_1m_2/(m_1+m_2)$ in a static Schwarzschild background of mass $m_1+m_2$.\\[10pt]
At the third post-Minkowksian order the result gets more interesting as
the result deviated from the one of a test particle in a Schwarzschild
background and gravitational radiation enter in a new and interesting
way.
At the third post-Minkowskian order the scattering angle read

\begin{multline}\label{e:Chi3PM}
{  \chi}_{\rm 3PM}= \frac{2 \left(64 \gamma
    ^6-120 \gamma ^4+60 \gamma ^2-5\right)}{3 \left(\gamma
    ^2-1\right)^{3\over2}}\left(G_Nm_1m_2\over J\right)^3\cr
+{8m_1m_2\sqrt{\gamma^2-1}\over 3\mathcal E_{\rm C.M.}^2}\Bigg(-\gamma(25+14\gamma^2)+\frac{3(3+12
  \gamma^2-4\gamma^4)\arccosh(\gamma)}{\sqrt{\gamma^2-1}}\Bigg)
\left(G_Nm_1m_2\over J\right)^3\cr
+{1\over
  (4(\gamma^2-1))^{4-D\over2}}\left(-\frac{11}{3}+\frac{d}{d\gamma}
\left(\frac{(2\gamma^2-1)\arccosh(\gamma)}{\sqrt{\gamma^2-1}}
\right) \right)\cr
\times  {4m_1m_2(2\gamma^2-1)^2\over
    \mathcal E^2_{\rm C.M.}}\left(G_Mm_1m_2\over J\right)^3 +\mathcal O(4-D)\,.
\end{multline}
The term multiplying $ (4(\gamma^2-1))^{4-D\over2}$  is what can be viewed as radiation-reaction terms
in~\cite{Damour:2020tta,DiVecchia:2021ndb,Herrmann:2021tct,Bjerrum-Bohr:2021din}.  Because of infrared singularities, the scattering amplitude has
been dimensionally regularized at intermediate stages but the final
classical result is infrared finite in the limit $D \to 4$.
The radiation-reactions contributions have also been derived using
different amplitude-based methods: (1) High-energy
scattering~\cite{DiVecchia:2020ymx,DiVecchia:2021ndb}, (2) Linear
response to the angular
momentum~\cite{Damour:2020tta,Veneziano:2022zwh,Manohar:2022dea}, (3)
Reverse unitarity and the KMOC
formalism~\cite{Herrmann:2021lqe,Kosower:2018adc,
  Mougiakakos:2021ckm,Riva:2021vnj}.

At the next orders in the post-Minkowskian expansion, one can
na\"\i vely derive a scattering angle from the amplitudes contributions
derived in section~\ref{sec:4PM} by taking a Fourier transform with
respect to the momentum transfert
   \begin{align}  \mathcal M_{\rm 4PM}(\gamma,|b|)&:=\int_{\mathbb R^2} e^{i\underline q\cdot b}
    {\mathcal M_{\rm 4PM}(\gamma,\underline q^2)\over 4m_1m_2\sqrt{\gamma^2-1}}
                                                    {d^{2}\underline q\over (2\pi)^{2}}\\
\nonumber                                                    &= -{G_N^4 m_1^2m_2^2(m_1+m_2)^3\over
      2\pi b^3}\left(c^{\rm probe}(\gamma)+{m_1m_2\over (m_1+m_2)^2}
          c^{\rm next-to-probe}(\gamma)\right)
  \end{align}
  and derive a scattering angle by differentiating with respect the
  center-of-mass angular momentum
  $\mathcal J_{\rm C.M.}=m_1m_2\sqrt{\gamma^2-1} b/\mathcal E_{\rm C.M.}$ to get
  \begin{align}
    \chi_{\rm 4PM}(\gamma)&=-{\partial
                            {\mathcal  M}_{\rm
                            4PM}(\gamma,J)\over\partial \mathcal
                            J_{\rm C.M.}}\\
\nonumber                            &=\left(G_N
 m_1m_2\over \mathcal J_{\rm C.M.}\right)^4  {(\gamma^2-1) (m_1+m_2)^3\over
                                       \mathcal E_{\rm C.M.}^3} \left(c^{\rm probe}(\gamma)+{m_1m_2\over(m_1+m_2)^2}
          c^{\rm next-to-probe}(\gamma)\right).
      \end{align}
      The scattering angle is decomposed into a conservative part
      $\chi_{\rm 4PM}$ and corrections from the
radiation-reaction
\begin{equation}
  \chi_{\rm 4PM}= \chi^{\rm cons}_{\rm 4PM}+ \delta^{\rm rr} \chi^{\rm
    rel}_{\rm 4PM}  .
\end{equation}
From the  probe in~\eqref{e:Probe} and potential-potential and
radiation-radiation sectors of the next-to-probe
piece of the amplitude in~\eqref{e:nextotprobePRRR}, we obtain the
conservative part of the angle at the fourth post-Minkowskian order
\begin{equation}
  \chi^{\rm cons}_{\rm 4PM}= -{3\over8\pi}
  {(\gamma^2-1)(m_1+m_2)^3\over \mathcal E_{\rm C.M.}^3} \left(G_N m_1m_2\over
   \mathcal J_{\rm C.M.}\right)^4\,  \left(c^{\rm probe}(\gamma)+{m_1m_2\over(m_1+m_2)^2}  \widehat{\mathcal M}^{\rm
    PR}_{3}(\gamma)\right),
\end{equation}
in agreement with the results of~\cite{Bern:2021yeh, Bini:2021gat,Dlapa:2021npj,Dlapa:2022lmu}

From the potential-radiation sector in~\eqref{e:nextotprobePR}
we obtain, using the notations of~\cite{Bini:2021gat}
 \begin{equation}
\delta^{\rm rr} \chi^{\rm rel}_{\rm 4PM}= -{3\over8\pi} \left(G_N m_1m_2\over
  \mathcal  J_{\rm C.M.}\right)^4 \left(2\chi_2^{\rm
      cons} J_2-h E_3{d\chi_1^{\rm cons}\over d\gamma}\right).
  \end{equation}
This expression reproduces the equation~(7.9) of~\cite{Bini:2021gat}
except for the $J_3$ piece which is affected by
the radiation loss~\cite{Bini:2021gat,Manohar:2022dea}. The radiation
changes the relation between the amplitude and the scattering
angle.   It has been argued in~\cite{DiVecchia:2022piu} that the eikonal connection between 
the momentum $q$-space analysis and the $b$-space results have to be 
corrected in order to include the angular momentum change from the
radiation. The consequences of this, for the scattering amplitude approach we have presented here, 
is still at the time of writing being researched.
\\[5pt]

\section{Conclusion}
We deliver in this review an account of several diverse and exciting developments that we expect will drive forward the field of gravitational physics in the future. An essential starting point for the analysis is the utilization of the framework of using effective field theory for perturbative gravitational interactions. This narrative guides naturally how to arrive at a low-energy quantum extension of General Relativity and is an ideal starting point for most contemporary phenomenological explorations of gravitational interactions. It is also an essential component in precision data analysis where gravitational scattering amplitudes are computed at high perturbative orders. Such computations are demonstrated to be beneficial as a complement to numerical General Relativity in the inspiral region of a binary black hole merger at relativistic velocities. Thus gravitational scattering amplitude technology is vital anywhere where computational precision is needed for an accurate analysis. We anticipate that advancement in this field to persist along the lines of the research summarized here. A current bottleneck for such computations arises in the determination of an ansatz for the scattering amplitude at high perturbative orders. We also envision further improvements in integrand reconstructions at a high multiplicity of exchanged gravitons in interactions. The current state of the art in computation is fifth post-Minkowskian order and we envision progress in the coming years. \\[5pt]
Another area of attraction is the computation of two-to-two scattering amplitude with spinning massive particles. Since black holes have non-trivial classical spins, the consequences of having a non-trivial spin can have important effects on the analysis. Spin can be accounted for in current numerical relativity computations, but an interesting idea is to employ perturbative amplitudes with spin to complement effects. Again the inspiral region of mergers is intriguing to examine since precision is indispensable. Here investigations are currently behind the state-of-the-art progress for high perturbative probes, but there are multiple attractive paths to track for additional progress. Since it is outside the extent of this review we depart such deliberations and leave them to forthcoming studies.\\[5pt]
Furnished with the effective field theory framework for a low energy theory of gravity it is also interesting to examine the potential consequences of high derivative terms in the effective Lagrangian. The higher derivative terms in the Lagrangian signal departure from Einstein's gravity, however since curvature is small in most instances, it is (very) challenging to realize areas where we would be capable of affirming deviation from Einstein's theory of relativity. Present bounds for verifiable cosmological deviations are huge. Nevertheless, since an effective field appears to be the natural framework for a likely theory of quantum gravity we should scrutinize any manifestation of the existence of higher derivative terms in the gravitational action. \\[5pt]
Increased resolution of observed gravitational wave signals stimulates quests for new physics and this opens a very exciting chapter of physics in the years to come. With our new theoretical laboratory for tests of low-energy phenomenological extensions of Einstein's theory of gravity, we envision providing potential for observation-driven astrophysical discoveries. \\[5pt]
%


\section*{Acknowledgments}
The research of P.V. has received funding from the ANR grant ``SMAGP'' ANR-20-CE40-0026-01. The work of N.E.J.B.-B. was supported by DFF grant 1026-00077B and in part by the Carlsberg Foundation.\\[15pt]

\end{document}